\documentclass[manuscript]{acmart}
\AtBeginDocument{%
  }

\setcopyright{acmlicensed}
\copyrightyear{2018}
\acmYear{2018}
\acmDOI{XXXXXXX.XXXXXXX}

\acmJournal{TOSEM}
\acmVolume{37}
\acmNumber{4}
\acmArticle{111}
\acmMonth{8}

\usepackage{amsmath,amssymb,amsfonts}
\usepackage{algorithmic}
\usepackage{graphicx}
\usepackage{textcomp}
\usepackage{colortbl}
\usepackage{geometry}
\usepackage{multirow}
\usepackage{url}
\usepackage{colortbl}
\usepackage[table, dvipsnames]{xcolor}
\usepackage{listings}
\usepackage{mathabx}
\usepackage{framed}

\theoremstyle{plain}

\theoremstyle{definition}

\theoremstyle{remark}

\begin{document}

\title{Context-Augmented Code Generation Using Programming Knowledge Graphs}


\author{Shahd Seddik}
\email{shahd.seddik@ubc.ca}
\orcid{0009-0005-2554-0083}
\author{Fahd Seddik}
\email{fahd.seddik@ubc.ca}
\orcid{0009-0000-9365-1096}
\author{Iman Saberi}
\email{iman.saberi@ubc.ca}
\orcid{0000-0003-2310-1376}
\author{Fatemeh Fard}
\email{fatemeh.fard@ubc.ca}
\orcid{0000-0002-4505-6257}
\affiliation{%
  \institution{University of British Columbia}
  \city{Kelowna}
  \country{Canada}
}

\author{Minh Hieu Huynh}
\email{minhhieuh@student.unimelb.edu.au}
\orcid{0000-0003-2310-1376}
\author{Patanamon Thongtanunam}
\email{patanamon.t@unimelb.edu.au}
\orcid{0000-0001-6328-8839}
\affiliation{%
  \institution{University of Melbourne}
  \city{Victoria}
  \country{Australia}
}

\renewcommand{\shortauthors}{Seddik et al.}

\begin{abstract}
  Large Language Models (LLMs) excel at code generation but struggle with complex problems. Retrieval-Augmented Generation (RAG) mitigates this issue by integrating external knowledge, yet retrieval models often miss relevant context, and generation models hallucinate with irrelevant data. We propose Programming Knowledge Graph (PKG) for semantic representation and fine-grained retrieval of code and text. Our approach enhances retrieval precision through tree pruning and mitigates hallucinations via a re-ranking mechanism that integrates non-RAG solutions. Structuring external data into finer-grained nodes improves retrieval granularity. Evaluations on HumanEval and MBPP show up to 20\% pass@1 accuracy gains and a 34\% improvement over baselines on MBPP. 
    Our findings demonstrate that our proposed PKG approach along with re-ranker effectively address complex problems while maintaining minimal negative impact on solutions that are already correct without RAG. The replication package is published at \url{https://github.com/iamshahd/ProgrammingKnowledgeGraph}.
    
\end{abstract}

\ccsdesc[500]{Computing methodologies~Natural language processing}
\ccsdesc[300]{Software and its engineering}
\keywords{Programming Knowledge Graph, Retrieval Augmented Generation, LLM Code Generation}


\maketitle

\section{Introduction}

Large Language Models (LLMs) have enabled natural-language-to-code generation on a wide range of programming tasks \cite{chen2021evaluating, huang2023agentcoder,roziere2023code}. However, functional correctness often depends on external programming knowledge that is not consistently stored in model parameters, including API usage conventions, corner cases, and idiomatic patterns. Consequently, Retrieval-Augmented Generation (RAG) has become a practical mechanism for grounding code generation in external sources such as library documentation, tutorials, and code repositories \cite{zhoudocprompting, zhang2023repocoder}. Recent large-scale evaluations further show that end-to-end gains depend jointly on retrieval quality and the generator's ability to use retrieved context effectively \cite{wang-etal-2025-coderag}.

At the same time, code-oriented RAG remains brittle. Prior work has shown that retrieved context can be redundant, partially relevant, or misleading, and that models may under-utilize or be distracted by long contexts \cite{wang-etal-2025-coderag}. More broadly, analyses of context utilization caution that performance measured under simplified or synthetic assumptions can overestimate how effectively models use retrieved context under realistic retrieval distributions \cite{hagstrom-etal-2025-reality}. Recent RAG pipelines for code have explored diverse retrieval strategies and knowledge bases \cite{su-etal-2024-evor}, yet systematic evidence suggests two recurring bottlenecks: accurately identifying helpful evidence for a given programming query, and operating under limited context budgets where irrelevant retrieval can directly harm generation \cite{wang-etal-2025-coderag}.

A central reason these bottlenecks persist is that actionable programming knowledge is heterogeneous. Developers draw from code-centric artifacts (e.g., implementations, utilities, reusable patterns) as well as text-centric artifacts (e.g., tutorial narratives, API documentation, and Q\&A explanations) \cite{wang-etal-2025-coderag}. This heterogeneity is not incidental; software maintenance evidence indicates that code and natural-language documentation often co-evolve, implying that text encodes complementary signals rather than redundant descriptions of code \cite{pai2025codocbench}. As a result, the representation used for retrieval must cope with fundamentally different structures across artifacts.

This raises an under-specified design problem. For code generation, what should be retrieved, and in what representation, to preserve useful knowledge while minimizing noise? Empirical studies show that different retrieved information types have qualitatively different values. Context-relevant information can help, while superficially similar snippets can introduce noise and degrade results \cite{gu2025retrieve}. These findings motivate knowledge representations that are explicitly designed to separate, organize, and filter heterogeneous knowledge rather than treating all data sources as flat nearest-neighbor chunks \cite{jiang2025ras}.

Recent work supports the principle that structure can materially change retrieval behavior and downstream generation \cite{sun-etal-2025-enhancing-retrieval}. For code-centric corpora, structure-aware chunking based on Abstract Syntax Trees (ASTs) yields more coherent retrievable units and improves retrieval-augmented code generation relative to line-based heuristics \cite{zhang-etal-2025-cast}. For long or redundant sources, tree-structured search improves over flat ranking \cite{sun-etal-2025-enhancing-retrieval}. In addition, work on granularity selection shows that the size of retrieval units induces an inherent precision--recall trade-off that can substantially affect downstream performance \cite{zhong-etal-2025-mix}. Collectively, these results motivate studying representations that (i) align retrieval units with structure and (ii) provide explicit control over how context is assembled.

In this work, we present \emph{\textbf{Programming Knowledge Graphs (PKGs)}}, a novel representation for RAG, using two stand-alone variants to isolate the role of knowledge modality and representation, being code-centric PKG and text-centric PKG.
Through PKG, we address the two challenges of  accurately identifying and retrieving helpful documents for code generation. 
The \emph{code-centric PKG} is constructed by parsing source code into an AST-derived hierarchy and then constructing a directed structure that mirrors syntactic containment. A function node expands to a sequence of block nodes, each of which may expand to child block nodes (and so on), yielding a multi-level tree rooted at the function.
This construction aligns retrieval units with syntactic boundaries and preserves self-contained code regions, consistent with evidence that AST-respecting segmentation produces more coherent retrieval units for retrieval-augmented code generation \cite{zhang-etal-2025-cast}.

A key advantage of this hierarchy is that it enables retrieval at multiple granularities.
We instantiate two retrieval settings: \emph{Func-PKG}, which retrieves at the function level (coarser units), and \emph{Block-PKG}, which retrieves at the block level (finer units).
These settings operationalize an inherent granularity trade-off: coarser units tend to increase recall but risk injecting irrelevant content, while finer units can improve precision but may omit surrounding context needed for correct synthesis.
Prior RAG work highlights that chunk granularity directly affects retrieval precision and recall and that selecting an appropriate granularity can significantly change downstream performance \cite{zhong-etal-2025-mix}.
We therefore evaluate both granularity settings to quantify this precision--recall trade-off in code-centric PKG retrieval and to determine when finer structure improves (or harms) end-to-end code generation.

The \emph{text-centric} PKG targets tutorial/documentation-style knowledge, where usefulness often depends on internal structure (titles, sections, explanations, examples) and on the ability to retrieve specific fields rather than entire passages \cite{jiang2025ras}.
We adopt a structured JSON representation as an explicit schema over heterogeneous tutorial fields, then construct a Directed Acyclic Graph (DAG) whose nodes correspond to JSON fields/sections and whose edges encode structural relations (e.g., example-to-explanation linkage). The intent is twofold: (i) normalize diverse textual artifacts into a parseable representation that supports field-level retrieval and assembly; and (ii) enable structure-aware traversal and composition, following the general retrieval-and-structuring perspective that hierarchical organization can improve controllability and reduce noise \cite{jiang2025ras,zhang2025graphragsurvey}.

Because structured expansion can still overgrow context windows, we incorporate \emph{tree pruning} as an explicit control mechanism before prompt construction. This design follows evidence that pruning irrelevant context can reduce computational overhead and mitigate noise propagation in retrieval-augmented generation \cite{chirkova2025provence}.
Finally, even with improved context, different approaches excel at solving distinct types of problems \cite{tao2025retrieval}. We therefore complement retrieval with \emph{post-generation reranking} (sample--then--select), where multiple candidate solutions from multiple methods (e.g., RAG and non-RAG approaches) are generated and then ranked to select the best. Recent work on code generation studies candidate reranking explicitly and reports improved success by selecting among diverse samples, supporting reranking as a practical method applied to retrieval \cite{sorokin2025reinfRerank,cao2025codersa}.
The reranker, therefore, can prioritize solutions generated without relying on RAG-based content, reducing the influence of erroneous data when the retrieved content introduces hallucinations into the output.

We compare NoRAG, sparse retrieval (e.g., BM25), dense retrieval, and PKG-based retrieval under both code-centric and text-centric variants, and we analyze when and why performance changes via topic-level breakdowns, error-type shifts, and cost trade-offs.
We evaluate our method using HumanEval \citep{chen2021evaluating} and MBPP \citep{austin2021mbpp}. Our approach improves the pass@1 accuracy across all baseline models on both the HumanEval \citep{chen2021evaluating} and MBPP \citep{austin2021mbpp} benchmarks by up to 20\% compared to the NoRAG method. In comparison to sparse and dense retrieval, 
our method demonstrates up to an 8\% increase in accuracy on HumanEval and up to a 34\% improvement on MBPP.
Error analysis on the MBPP dataset, which contains more and complex problems, reveals that Assertion errors are reduced significantly, though Name errors are introduced. 
Additionally, topic analysis on MBPP demonstrate the difficulty of solving some problems e.g., string manipulation when using RAG based on PKG.

Our main contributions can be summarized as follows:
\begin{enumerate}
  \item \textbf{A novel representation of code-centric and text-centric corpora in the form of a PKG for retrieval-augmented code generation.}
  We position PKGs as a structured knowledge representation for retrieving context during code generation.

  \item \textbf{Code-centric PKG construction with an explicit granularity comparison.}
  We construct a code-centric PKG by parsing code into an AST-derived hierarchy and materializing a containment structure (function $\rightarrow$ block $\rightarrow$ child block $\rightarrow \cdots$). We evaluate retrieval at two granularities---\emph{Func-PKG} (function-level) and \emph{Block-PKG} (block-level).

  \item \textbf{Text-centric PKG construction via structured JSON and DAG extraction.}
  We construct a text-centric PKG by converting tutorial/documentation-style text into structured JSON and extracting a DAG, enabling field-level retrieval and structured context assembly.

  \item \textbf{Structure-based context control through pruning.}
  We incorporate pruning over retrieved graph structure prior to prompt construction to control context growth and reduce inclusion of low-utility branches.

  \item \textbf{Post-generation reranking of solution candidates as an orthogonal selection mechanism.}
  We apply reranking over multiple generated candidate solutions to improve final correctness.

\end{enumerate}

The rest of the paper is organized as follows. Section \ref{sec:related-work} discusses related works. Section \ref{sec:methodology} explains the methodology. Sections \ref{sec:experimental-setup} and \ref{sec:results} present the experimental setup and results, respectively. Section \ref{sec:discussions} provides a discussion of the findings, error analysis, cost trade-off, and implications. Section~\ref{sec:threats} discusses the threats to the validity of this research.
Finally, Section \ref{sec:conclusion} concludes the paper.

\section{Related Work}
\label{sec:related-work}
In this section, we provide an overview of the related literature and position the differences of our work with previous studies. 

\subsection{Retrieval Augmented Generation}

RAG approaches are widely explored in general text generation \citep{guu2020retrieval, lewis2020retrieval, jiang2023active, gao2023retrieval}.
RAG continues to evolve along three mutually reinforcing directions: (i) structure- and graph-aware retrieval that exploits document- or event-level topology; (ii) semantic pre-processing that produces more coherent retrieval units; and (iii) architectural optimizations that trade off latency, memory and fidelity. Our approach is best characterized as structure-aware (graph-based) RAG, since it explicitly represents code and tutorial content as a PKG and retrieves connected code blocks and path-value nodes from that graph. It also incorporates semantic pre-processing (AST/JSON chunking) and an advanced re-ranking stage, but it does not target architectural/decoding speedups.  

\textbf{Structure- and graph-aware retrieval.} Recent work shows clear gains from retrieving connected, topology-aware units rather than isolated passages. StructRAG builds document-level scholarly KGs and reports modest semantic-accuracy gains alongside larger improvements in lexical diversity and readability (while flagging KG construction cost/scalability) \cite{jia2025structrag}; KG\textsuperscript{2}RAG augments seed retrieval with KG-guided expansion and KG-driven linearization (MST+DFS) and can use either external KGs or LLM-extracted triplets, so its effectiveness depends on KG coverage and linking quality. GRAG formalizes \emph{textual subgraph} retrieval and fuses text+topology views for generation, showing strong gains on graph-centric multi-hop tasks but limited cross-domain validation \cite{hu2025grag}. Domain variants (GeoGraphRAG, EventRAG) demonstrate clear task-specific benefits (geospatial operators, temporal/event fusion) while illustrating portability and modality limits \cite{liang2025geographrag,yang2025eventrag}.

\textbf{Semantic pre-processing / chunking.} Methods that produce semantically coherent retrieval units improve downstream precision: SemRAG emphasizes semantic chunking and KG-style organization and documents the granularity trade-off (precision vs.\ recall/coverage) \cite{zhong2025semrag}; KG\textsuperscript{2}RAG and related pipelines supplement chunking with triplet extraction or entity-linking to form richer contexts, but these steps add engineering cost and propagate KG/linking errors when automated \cite{zhu2025kg2rag}.

\textbf{Architectural / decoding optimizations.} A separate line targets latency and memory: REFRAG compresses and selectively expands retrieved chunks during decoding to exploit block-sparse attention, reporting large improvements in time-to-first-token ($\approx 30\times$) and competitive perplexity, though it requires additional encoder/projector/policy training and careful validation for fidelity and compatibility with deployed LLMs \cite{lin2025refrag}.

\subsection{RAG for Code Generation}


RAG's use in code-related tasks remains underexplored \citep{wang2024coderag}. Previous studies, like \cite{parvez2021retrieval}, focused on smaller models like CodeBERT and GraphCodeBERT for tasks like code summarization and generation, often fine-tuning the retriever. In contrast, our approach applies RAG during inference without fine-tuning. 
DocPrompting \cite{zhou2022docprompting} is designed to support code generation across multiple programming languages and relies on retrieving information from a documentation pool. It has also been tested on smaller-scale models. 
RepoCoder \cite{zhang2023repocoder} is specifically designed for repository-level code completion, leveraging an iterative retrieval-generation pipeline to refine the generated code. 
HippoRAG \cite{gutierrez2024hipporag} is another approach that introduces a neurobiologically inspired approach for integrating long-term memory into large language models, primarily focusing on natural language tasks that require knowledge aggregation from multiple graph-based sources. While both approaches leverage structured knowledge retrieval, our method is tailored to programming-related tasks, emphasizing the seamless integration of documentation and code to improve the quality of generated outputs. 

Recent literature highlights the efficacy of incorporating structural information into RAG frameworks for code. In the context of code completion, RepoCoder \cite{zhang2023repocoder} establishes a baseline for structure-aware RAG, utilizing a multi-granularity framework that streamlines repository-level completion and achieves over 10\% improvement over non-structural baselines. Subsequent works have sought to capture more intricate dependencies through Knowledge Graphs (KGs). For example, KGCompass \cite{yang2025enhancingrepositorylevelsoftwarerepair} constructs a repository-aware KG to map issues to specific code entities, using entity path tracing to narrow the search space for software repair. Similarly, Prometheus \cite{chen2025prometheusunifiedknowledgegraphs} unifies files, ASTs, and natural language into a heterogeneous graph with typed edges, demonstrating effectiveness across seven programming languages. More recent approaches integrate these structural representations with agentic workflows. CodexGraph \cite{liu2025codexgraph} interfaces LLMs with graph databases to support complex, multi-task operations, while LingmaAgent \cite{ma2025alibabalingmaagentimprovingautomated} employs a Monte Carlo tree search strategy over a condensed KG to enable top-down repository exploration. Finally, RepoGraph \cite{ouyang2025repographenhancingaisoftware} provides a plug-in module, facilitating interactive navigation of repository-level structure in AI-driven engineering solutions.

\textbf{Differences of our work with the existing literature.}
While \citep{wang2024coderag} explores LLMs and Code-LLMs across data sources, they note issues with retrievers and limited model context. Our method improves knowledge representation, enabling more accurate retrieval and reducing hallucinations by prompting models with only relevant content.
A fundamental difference of PKG with \cite{zhou2022docprompting} lies in the retrieval mechanism. Our method decomposes content into fine-grained semantic nodes within the PKG, integrating both documentation and code. 
In contrast to RepoCoder \cite{zhang2023repocoder}, our PKG-based approach is adaptable to a broader range of code-centric and text-centric datasets, making it more versatile in various software development scenarios and can be a complement to RepoCoder. Our retrieval mechanism prioritizes extracting the most relevant, fine-grained content from the knowledge base. Unlike RepoCoder's iterative retrieval strategy, which refines results over multiple retrieval cycles, our method ensures that only the most pertinent information is retrieved from the outset, leading to more efficient and precise code generation. 

Despite these advancements, the effective utilization of structural granularity (specifically the trade-off between coarse-grained and fine-grained retrieval units) remains under-explored in current RAG frameworks. While approaches like RepoCoder \cite{zhang2023repocoder} and DocPrompting \cite{zhou2022docprompting} leverage structure, they typically rely on iterative retrieval cycles or coarse document pooling, which can introduce irrelevant context or obscure precise semantic boundaries. Furthermore, many existing graph-based methods \cite{liu2025codexgraph, ma2025alibabalingmaagentimprovingautomated} focus on repository-wide navigation via complex agents rather than optimizing the immediate representation of knowledge for generation. In contrast, our work introduces a novel inference-time framework that creates distinct hierarchical representations for code (via AST-based containment) and documentation (via JSON-based DAGs). Unlike prior methods that require fine-tuning or undefined granularity, our approach explicitly operationalizes the precision-recall trade-off through variable retrieval units (e.g., \emph{Block-PKG} vs. \emph{Func-PKG}) and employs structural pruning to actively minimize noise within the context window.



\section{Methodology}
\label{sec:methodology}
Our method consists of three stages: (i) PKG construction, (ii) retrieval from the PKG to form an augmented prompt, and (iii) candidate selection via reranking. Figure~\ref{fig:PKG-generation} summarizes the construction pipeline and Figure~\ref{fig:PKG-retrieval} summarizes retrieval.

\subsection{Programming Knowledge Graph (PKG) Construction}
\label{sec:pkg-generation}

\paragraph{Data sources.}
We build two PKGs from (1) a code-centric corpus derived from PythonAlpaca~\citep{nic2024pyalpaca} and (2) a text-centric corpus derived from a Python-focused subset of Tutorials~\citep{wang2024coderag} (Step 1 in Figure~\ref{fig:PKG-generation}). Let $\mathcal{D}^{\texttt{code}}$ denote the code-centric dataset and $\mathcal{D}^{\texttt{text}}$ denote the text-centric dataset.

\paragraph{Graph schema.}

The PKG stores each artifact in multiple, consistent granularities so that retrieval can trade off context breadth and specificity. Each code artifact is represented at (i) function granularity (the full implementation) and (ii) block granularity (syntactic sub-structures). Each tutorial artifact is represented as a hierarchy, where leaves correspond to primitive values addressed by their JSON paths.

We model the PKG as a typed, directed graph $G=(V,E,\tau,\phi)$ where $V$ is a set of nodes, $E \subseteq V \times V$ is a set of directed edges, $\tau:V \rightarrow \mathcal{T}$ assigns a node type, and $\phi:V \rightarrow \Sigma^{*}$ maps each node to a textual payload (code or text) that is embedded for retrieval (Section~\ref{sec:embedding_storage}). Edges implement a refinement relation between granularities: a coarse unit refines into finer units (function $\rightarrow$ blocks, JSON section $\rightarrow$ nested leaves). This refinement relation is acyclic because syntactic nesting and JSON nesting are acyclic. A key property is that any retrieved fine-grained unit has a well-defined projection back to its parent coarse unit, enabling the system to return either precise snippets or broader context depending on the retrieval mode (Section~\ref{sec:retrieval}).

For completeness, Appendix~\ref{app:pkg-schema} enumerates the concrete node and edge types used in our implementation, including identifier-to-content links for code and containment links for both code and JSON-derived structures.

\subsubsection{Code-centric extraction: functions and block hierarchy}
\label{sec:code_extraction}

For each example in $\mathcal{D}^{\texttt{code}}$, we extract Python functions from the code field(s) and parse them using a Python AST parser. For a function $F$, let $\mathcal{B}(F)$ denote the set of syntactic blocks within $F$ (for example, \texttt{if}, \texttt{for}, \texttt{while}, \texttt{try}, \texttt{with}, and function-local compound statements). Each block $b \in \mathcal{B}(F)$ is represented by its textual payload $\phi(b)$.

For each function $F$, we create a subgraph $G_F=(V_F,E_F,\tau,\phi)$ consisting of one \textsc{Name} node, one \textsc{Impl} node, and one \textsc{Block} node per extracted block:
\begin{equation}
V_F = \{ v_{\textsc{Name}}^{F}, v_{\textsc{Impl}}^{F} \} \cup \{ v_{\textsc{Block}}^{F,b} : b \in \mathcal{B}(F) \}.
\end{equation}
Edges encode containment and nesting:
\begin{align}
E_F \;=&\; \{(v_{\textsc{Name}}^{F}, v_{\textsc{Impl}}^{F})\} \;\cup\; \{(v_{\textsc{Impl}}^{F}, v_{\textsc{Block}}^{F,b}) : b \in \mathcal{B}(F)\} \nonumber \\
&\;\cup\; \{(v_{\textsc{Block}}^{F,b_p}, v_{\textsc{Block}}^{F,b_c}) : b_c \text{ is a direct syntactic child of } b_p\}.
\label{eq:code_edges}
\end{align}
The \textsc{parent} edges form a directed acyclic graph (DAG) because syntactic nesting is acyclic. This step is shown in Step 3 of Figure~\ref{fig:PKG-generation}.

\subsubsection{Text-centric extraction: JSON path-value graph}
\label{sec:json_extraction}

For each document $x \in \mathcal{D}^{\texttt{text}}$, we produce a JSON representation $J(x)$ using a constrained prompting procedure and validate the output by JSON parsing. We model each JSON as a rooted tree whose nodes are keys and whose edges follow JSON nesting.

A JSON value is either a \emph{primitive} (string, number, boolean), an \emph{object} (a finite map from keys to values), or an \emph{array} (an ordered list of values). We view an object $J$ as a rooted tree where:
(i) the \emph{root} consists of the top-level keys in the JSON file, and
(ii) a key has children when its associated value is itself an object (or an array whose elements are objects).
Thus, nested JSON objects induce parent-child relationships between keys, and leaves correspond to keys whose values are primitives.

For any key reachable in this tree, its \emph{path} is the sequence of keys from a top-level key to that key. For a key $k$, we write its path as
$p(k) = (k_0, k_1, \ldots, k_\ell)$, where $k_0$ is a top-level key and $k_\ell = k$. We use a string form $\mathrm{Join}(p)$ (for example, \texttt{k0/k1/.../k$\ell$}) as a unique identifier.

We create \textsc{PathValue} nodes only for \emph{leaf} paths whose associated value is primitive. Let $\mathrm{Leaves}(J)$ denote the set of leaf keys in $J$, and let $val(k)$ be the value associated with key $k$. For each $k \in \mathrm{Leaves}(J)$, we create a node $v^{k}$ of type \textsc{PathValue} with payload:
\begin{equation}
\phi(v^{k}) = \mathrm{Serialize}\!\left(\mathrm{Join}(p(k)),\, val(k)\right),
\end{equation}
where $\mathrm{Serialize}$ produces a short textual form such as \texttt{path = ... ; value = ...}.

To preserve hierarchy, we materialize internal path nodes and add directed edges from each parent path to its child path. Since JSON nesting is tree-structured, the resulting directed graph is acyclic. This step is illustrated in Step 3 of Figure~\ref{fig:PKG-generation}.

\subsubsection{Embedding and storage}
\label{sec:embedding_storage}

\begin{figure}
    \centering
    \includegraphics[width=0.8\columnwidth]{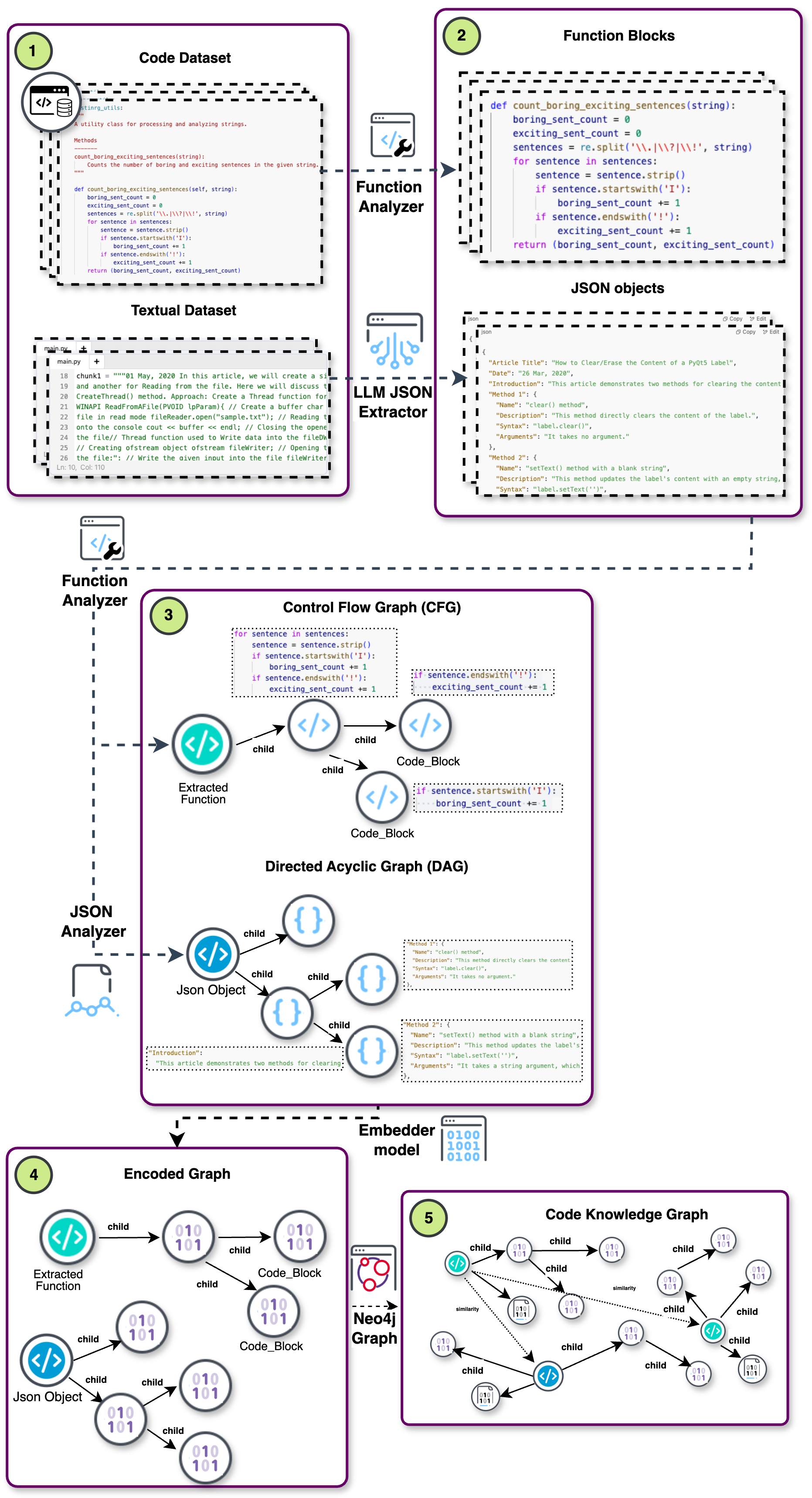}
    \caption{The overview of generating PKG}
    \label{fig:PKG-generation}
    \Description{The overview of generating PKG}
\end{figure}

We embed each node payload for semantic retrieval. Let $\mathcal{E}:\Sigma^{*}\rightarrow\mathbb{R}^{d}$ be an embedding model. Each node $v \in V$ stores an embedding vector:
\begin{equation}
\mathbf{z}_v = \mathcal{E}(\phi(v)).
\end{equation}
As shown in step 4 of Figure~\ref{fig:PKG-generation}, we store $G$ and $\{\mathbf{z}_v\}_{v \in V}$ in a graph database with a vector index to support approximate nearest-neighbor search.

\subsection{Retrieval from PKG}
\label{sec:retrieval}

\begin{figure*}[htbp]
    \centering
    \includegraphics[width=1\textwidth]{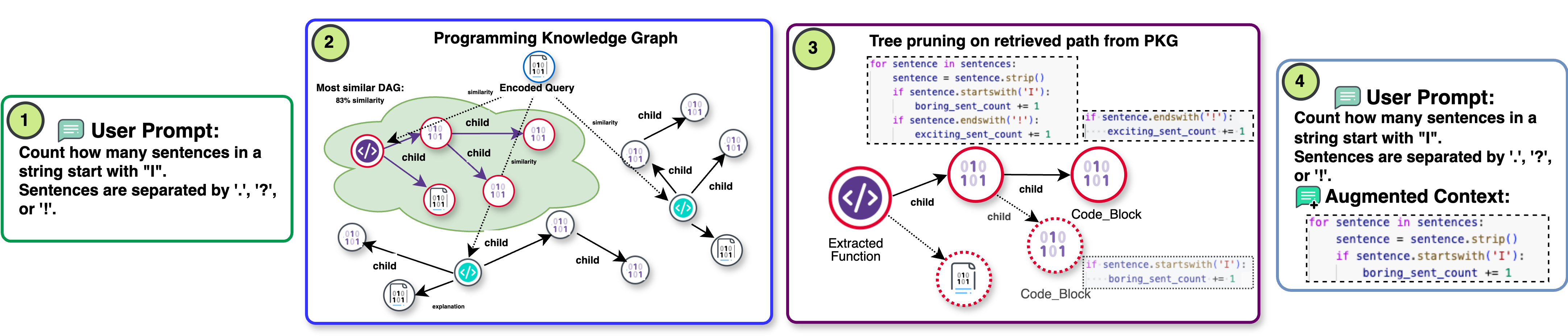}
    \caption{Overview of the retrieval process from PKG}
    \label{fig:PKG-retrieval}
    \Description{Overview of the retrieval process from PKG}
\end{figure*}

As in step 1 in Figure~\ref{fig:PKG-retrieval}, given a user query $q \in \Sigma^{*}$, we compute its embedding $\mathbf{z}_q = \mathcal{E}(q)$ and retrieve relevant nodes under three retrieval modes: block-wise (results will be labeled \textsc{Block-PKG}), function-wise (results will be labeled \textsc{Func-PKG}), and tutorial path-value (results will be labeled \textsc{JSON-PKG}). Let $V_t = \{v \in V : \tau(v)=t\}$ be the set of nodes of type $t$.
For a retrieval mode $t \in \{\textsc{Block},\textsc{Impl},\textsc{PathValue}\}$, we retrieve the top node based on the following objective (Steps 2 and 3 in Figure~\ref{fig:PKG-retrieval}):
\begin{equation}
v^*(q) = \arg\max_{\substack{v \in V_t}} \mathrm{Sim}(q,v).
\label{eq:topk}
\end{equation}
where $\mathrm{Sim}(\cdot, \cdot)$ refers to the cosine similarity. In practice, we compute $v^*$ using a vector index.


\paragraph{Pruning of block subtrees.}
Given a retrieved code node $v^*$ (either a \textsc{Impl} node for function-wise retrieval or a \textsc{Block} node for block-wise retrieval), we optionally remove irrelevant branches from its associated syntactic containment DAG to better align the returned context with the query (Step 3 in Figure~\ref{fig:PKG-retrieval}).

Let $G_{v^*} = (V_{v^*}, E_{v^*})$ denote the DAG induced by $v^*$ and its descendant \textsc{Block} nodes under \textsc{parent} edges. We define a \emph{branch} as a child-subtree rooted at a direct child of $v^*$ in $G_{v^*}$. Let $\mathrm{Child}(v^*)$ be the set of direct children of $v^*$ in this DAG. For each $u \in \mathrm{Child}(v^*)$, let $T_u(v^*)$ be the subtree rooted at $u$ (including all descendants of $u$). We define the pruned graph obtained by removing that branch as
\begin{equation}
G_{v^*}^{-u} \;=\; G_{v^*} \setminus T_u(v^*),
\end{equation}
where $G \setminus T$ removes the nodes in $T$ and all incident edges from $G$.

To select which branch to remove, we score each pruned graph by its cosine similarity to the query embedding. Concretely, we serialize the remaining graph content into text $\phi(G_{v^*}^{-u})$ (by concatenating the code payloads of its remaining nodes in a fixed order), embed it using the same embedder $\mathcal{E}$, and compute its cosine similarity. The selected pruned graph is
\begin{equation}
G_{\text{pruned}} \;=\; \arg\max_{u \in \mathrm{Child}(v^*)} \mathrm{Sim}\!\left(q, G_{v^*}^{-u}\right).
\end{equation}
The retrieved context returned to the generator is then the serialized content $\phi(G_{\text{pruned}})$.

\paragraph{Query augmentation.}
Let $t \in \{\textsc{Block}, \textsc{Impl}, \textsc{PathValue}\}$ denote the active retrieval mode and let $v^*(q)$ be the single retrieved node from Equation~\ref{eq:topk}. We convert the retrieval result into a textual context block and append it to the original query (Step 4 in Figure~\ref{fig:PKG-retrieval}).

For tutorial retrieval ($t=\textsc{PathValue}$), the context is simply the payload of the retrieved node:
\begin{equation}
\mathcal{C}(q) \;=\; \phi\!\left(v^*(q)\right).
\end{equation}

For code retrieval ($t \in \{\textsc{Block}, \textsc{Impl}\}$), the context $\mathcal{C}(q)$ is giving by:
\begin{equation}
\mathcal{C}(q) \;=\; \phi(G_{\text{pruned}})
\end{equation}
where $G_{\text{pruned}}$ is defined by the pruning step applied to the induced DAG of $v^*(q)$.

Finally, we construct the augmented prompt by concatenating the user query with the retrieved context under a deterministic template:
\begin{equation}
q_{\textsc{aug}} \;=\; \mathrm{Format}\!\left(q,\, \mathcal{C}(q)\right),
\end{equation}
where $\mathrm{Format}$ inserts explicit delimiters and a brief instruction indicating that the model may use the provided context as reference. The augmented prompt $q_{\textsc{aug}}$ is then passed to the code generation model.

\subsection{Solution Reranking}
\label{sec:reranking}

As shown by recent research \cite{tao2025retrieval}, retrieval can both help and harm. We therefore select among multiple candidate solutions generated via different methods.

Let $M$ be the code generation model. For each task, we generate one candidate under each condition (NoRAG, BM25, PKG variants) using the same decoding settings. Let $C = \{c_1,\ldots,c_N\}$ be the union of all candidates.

We apply two deterministic filters:
\begin{itemize}
    \item \textbf{Syntactic validity:} $A(c)=1$ if $c$ parses as Python, else $0$.
    \item \textbf{Runtime sanity:} $R(c)=1$ if $c$ can be executed in a sandbox without raising exceptions, else $0$.
\end{itemize}
Let $C_A=\{c \in C : A(c)=1\}$ and $C_R=\{c \in C_A : R(c)=1\}$.

Among remaining candidates, we select using a similarity score between the query and candidate code:
\begin{equation}
c^{*} = \arg\max_{c \in C_R} \mathrm{Sim}(q, c),
\label{eq:rerank_select}
\end{equation}
where $\mathrm{Sim}(\cdot, \cdot)$ is the cosine similarity, and the candidate embedding is computed as $\mathcal{E}(c)$. We report oracle selection separately by choosing the first candidate that passes benchmark tests, which provides an upper bound on attainable gains from candidate diversity. Note that the reranker in Equation~\ref{eq:rerank_select} is intentionally simple and model-agnostic. Its role is to reduce retrieval-induced regressions by selecting candidates that remain aligned with the user query. We evaluate this reranker against both non-reranked settings and an oracle upper bound in Section~\ref{sec:results}.

\section{Experimental Setup}
\label{sec:experimental-setup}
\textbf{Retrieval Approaches:}
We utilized two retrieval methods based on a comparative analysis of various code retrieval models, as described by \cite{wang2024coderag}. For dense retrieval, we selected the Voyage-Code-2 model, recognized as one of the top-performing dense retrievers for code. Embeddings were obtained through API calls to this model. For sparse retrieval, we employed the BM25 algorithm, implemented using the $rank\_bm25$ Python library\footnote{https://pypi.org/project/rank-bm25/}, which exhibited the strongest performance among sparse retrieval techniques.

\textbf{Dataset and PKG Generation:}
We used the PythonAlpaca dataset \citep{nic2024pyalpaca} as a code-centric data source, which contains 143,000 general Python question-answer pairs. After preprocessing, we extracted 115,000 Python functions from the dataset. This extraction enabled us to construct a PKG comprising 425,058 nodes and 434,518 relations.

We also performed experiments with the Tutorials dataset \citep{wang2024coderag} as a text-centric data source, which contains 76,600 programming tutorial content. After converting them into json representations, pkg contains 288,583 path-value nodes and 287,936 relations. The graphs were generated using Neo4J version 5.20.0, optimized for handling large-scale graphs and supporting semantic search over the stored content.

\textbf{Code Generation Models:}
We conducted our experiments on four well-known Code-LLMs: CodeLlama-7B \citep{roziere2023codellama},CodeLlama-13B \citep{roziere2023codellama}, StarCoder2-7B \citep{lozhkov2024starcoder2}, and DeepSeek-Coder-7B \citep{zhu2024deepseek}. In addition, we tested Llama3.1-8B \citep{dubey2024llama3}, a general-purpose LLM that has demonstrated strong performance on code generation tasks. All experiments were conducted using a single A100 GPU.

\textbf{Evaluation Metric:}
To evaluate the accuracy of generated code, we used the pass@1 metric \citep{chen2021evaluating}. Due to resource constraints, we adopted a greedy decoding approach for the pass@1 evaluation, generating a single solution with a temperature setting of $t=0$ and a token limit of 512 $(max\_new\_tokens=512)$. Due to the deterministic nature of the scores, the results should be compared directly without requiring statistical tests. 

\textbf{Benchmarks:}
In this study, we aim to evaluate the general Python programming skills of both Code-LLMs and LLMs. To achieve this, we have selected the HumanEval dataset \citep{chen2021evaluating} and the MBPP benchmark \citep{austin2021mbpp}. These datasets are well-established in the literature and are widely used to assess both problem-solving and reasoning capabilities in Python programming.



\section{Results}
\label{sec:results}
In this section we carry out experiments to answer the following research questions. 

\begin{table*}[!ht]
\centering
\caption{Performance of code-centric retrieval-augmented code generation on HumanEval, reported as pass@1. Red cells indicate performance below NoRAG and green cells show scores above NoRAG, with color intensity reflecting significance. ``Ideal Reranker'' serves as an oracle upper bound for our proposed re-ranking stage.}
\vspace{2pt}
\renewcommand{\arraystretch}{1.3}
\scalebox{0.70}{
\begin{tabular}{l l cccccccc}
\toprule
\textbf{} & \textbf{Model} & \textbf{None} & \textbf{BM25} & \textbf{VoyageEmb} & \textbf{Func-BM25} & \textbf{Func-PKG} & \textbf{Block-PKG} & \textbf{Reranked} & \textbf{Ideal Reranker} \\ 
\midrule

& CodeLlama-7B      & 33\% \cellcolor{gray!0} & \cellcolor{OrangeRed!24}21\% & \cellcolor{ForestGreen!18}42\% & 33\% & \cellcolor{ForestGreen!10}38\% & \cellcolor{ForestGreen!14}40\% & \cellcolor{ForestGreen!26}\textbf{46\%} & \cellcolor{BlueViolet!20}56\% \\ 
& CodeLlama-13B     & 42\% \cellcolor{gray!0} & \cellcolor{OrangeRed!16}34\% & \cellcolor{ForestGreen!6}45\% & \cellcolor{ForestGreen!2}43\% & \cellcolor{ForestGreen!6}46\% & \cellcolor{ForestGreen!8}47\% & \cellcolor{ForestGreen!18}\textbf{51\%} & \cellcolor{BlueViolet!20}63\% \\ 
& Llama3.1-8B       & 55\% \cellcolor{gray!0} & \cellcolor{OrangeRed!42}34\% & \cellcolor{OrangeRed!10}50\% & \cellcolor{OrangeRed!2}54\% & 55\% & \cellcolor{ForestGreen!14}61\% & \cellcolor{ForestGreen!14}\textbf{66\%} & \cellcolor{BlueViolet!20}75\% \\ 
& StarCoder2-7B     & 45\% \cellcolor{gray!0} & \cellcolor{OrangeRed!8}41\% & \cellcolor{ForestGreen!16}53\% & \cellcolor{ForestGreen!24}57\% & \cellcolor{ForestGreen!22}56\% & \cellcolor{ForestGreen!28}59\% & \cellcolor{ForestGreen!36}\textbf{63\%} & \cellcolor{BlueViolet!20}72\% \\ 
& DeepSeek-Coder-7B & 70\% \cellcolor{gray!0} & \cellcolor{OrangeRed!52}44\% & \cellcolor{OrangeRed!20}60\% & \cellcolor{OrangeRed!18}62\% & \cellcolor{OrangeRed!2}69\% & \cellcolor{OrangeRed!2}68\% & \cellcolor{ForestGreen!6}\textbf{73\%} & \cellcolor{BlueViolet!20}83\% \\ 
& \textbf{Avg.}     & \textbf{49.0\%} & \textbf{34.8\%} & \textbf{50.0\%} & \textbf{49.8\%} & \textbf{52.8\%} & \textbf{55.0\%} & \textbf{59.8\%} & \textbf{69.8\%} \\



\bottomrule
\end{tabular}}
\label{table:different_methods_human-eval}
\end{table*}

\begin{table*}[t!]
\centering
\caption{Performance of code-centric retrieval-augmented code generation on MBPP, reported as pass@1. Rows are grouped into open-source and closed-source models. Red cells indicate accuracy below NoRAG, green cells indicate accuracy above, and color intensity reflects significance. ``Ideal Re-ranker'' serves as the upper bound for the proposed re-ranker method.}
\label{table:different_methods_mbpp}
\vspace{6pt}
\renewcommand{\arraystretch}{1.3}
\scalebox{0.70}{
\begin{tabular}{l l cccccccc}
\toprule
\textbf{} & \textbf{Model} & \textbf{None} & \textbf{BM25} & \textbf{VoyageEmb} & \textbf{Func-BM25} & \textbf{Func-PKG} & \textbf{Block-PKG} & \textbf{Reranked} & \textbf{Ideal Reranker} \\ 
\midrule

& CodeLlama-7B      & \cellcolor{gray!0} 38\% & \cellcolor{OrangeRed!22}27\% & \cellcolor{OrangeRed!12}32\% & \cellcolor{OrangeRed!22}27\% & \cellcolor{ForestGreen!12}44\% & \cellcolor{ForestGreen!16}46\% & \cellcolor{ForestGreen!40}\textbf{58\%} & \cellcolor{BlueViolet!20}60\% \\ 
& CodeLlama-13B     & \cellcolor{gray!0} 44\% & \cellcolor{OrangeRed!16}36\% & \cellcolor{OrangeRed!36}26\% & \cellcolor{OrangeRed!16}36\% & \cellcolor{OrangeRed!8}40\% & \cellcolor{ForestGreen!8}48\% & \cellcolor{ForestGreen!22}\textbf{55\%} & \cellcolor{BlueViolet!20}57\% \\ 
& Llama3.1-8B       & \cellcolor{gray!0} 43\% & \cellcolor{OrangeRed!10}38\% & \cellcolor{OrangeRed!4}41\% & \cellcolor{OrangeRed!4}41\% & \cellcolor{ForestGreen!6}46\% & \cellcolor{ForestGreen!12}49\% & \cellcolor{ForestGreen!40}\textbf{63\%} & \cellcolor{BlueViolet!20}66\% \\ 
& StarCoder2-7B     & \cellcolor{gray!0} 46\% & \cellcolor{OrangeRed!42}25\% & \cellcolor{OrangeRed!58}17\% & \cellcolor{OrangeRed!30}31\% & \cellcolor{OrangeRed!34}29\% & \cellcolor{ForestGreen!10}51\% & \cellcolor{ForestGreen!34}\textbf{62\%} & \cellcolor{BlueViolet!20}64\% \\ 
& DeepSeek-Coder-7B & \cellcolor{gray!0} 56\% & \cellcolor{OrangeRed!12}50\% & \cellcolor{OrangeRed!22}45\% & \cellcolor{OrangeRed!18}47\% & \cellcolor{OrangeRed!12}50\% & \cellcolor{OrangeRed!18}47\% & \cellcolor{ForestGreen!22}\textbf{65\%} & \cellcolor{BlueViolet!20}68\% \\ 
& \textbf{Avg.}     & \textbf{45.4\%} & \textbf{35.2\%} & \textbf{32.2\%} & \textbf{36.4\%} & \textbf{41.8\%} & \textbf{48.2\%} & \textbf{60.6\%} & \textbf{63.0\%} \\



\bottomrule
\end{tabular}
}
\end{table*}

\textbf{RQ1: Does code-centric PKG improve code generation?}

In this research question, we aim to explore the potential of leveraging graph-based retrieval-augmented methods on code-centric data source to improve code generation task. Specifically, we examine how relevant code retrieved from a PKG built on PythonAlpaca \cite{nic2024pyalpaca} can improve the performance of LLMs and Code-LLMs in generating accurate code.

Our method retrieves relevant code from the PKG and integrates it into the generation process (See section \ref{appendix:examples} in Appendix). We compare this approach to several baselines, detailed in Tables \ref{table:different_methods_human-eval} and \ref{table:different_methods_mbpp}, for HumanEval and MBPP benchmarks. The baselines include: 1) None: No retrieval-augmented generation, 2) BM25: Applied to the entire dataset without pre-processing, 3) VoyageEmb: Embeddings from question-answer pairs for retrieval, 4) Func-BM25: BM25 applied to function-extracted data, 5) Func-PKG: Semantic search over function-related nodes, 6) Block-PKG: Granular semantic search over code blocks for deeper context, 7) Reranked: Reranking of candidates from the retrieval methods 1-6, and 8) Ideal Reranker: Selecting the first solution that passes. This serves as an upper bound oracle simulating perfect reranking conditions, as the correct solution is shown to the model in the retreival process.

Tables \ref{table:different_methods_human-eval} and \ref{table:different_methods_mbpp} show that the effect of retrieval augmentation depends critically on (i) the retrieval unit and (ii) selection strategy. Naïve row-level retrieval is unstable: BM25 applied over unprocessed rows degrades performance for all models on both HumanEval and MBPP. For example, average pass@1 drops by 14.2 points on HumanEval and 10.2 points on MBPP relative to no retrieval. Dense retrieval over Q\&A pairs (``VoyageEmb'') is similarly inconsistent; on MBPP, dense retrieval underperforms the no-retrieval baseline by 13.2 points on average. These results indicate that retrieval is not inherently beneficial; when the retrieved context is noisy or weakly aligned to the target task, it can distract the model and reduce correctness.

In contrast, code-structured retrieval units improve robustness. When retrieval operates over extracted functions (Func-PKG) or smaller code blocks (Block-PKG), performance improves on average. Block-level retrieval yields the best average performance among non-reranked methods, consistent with the hypothesis that smaller units reduce irrelevant context and increase the density of actionable signals. However, block-level retrieval is not uniformly beneficial: DeepSeek-Coder-7B is negatively impacted by Block-PKG on both benchmarks. This aligns with observations from a related study by \citep{wang2024coderag}, where it exhibited similar behavior. Based on these findings, we hypothesize that DeepSeek-Coder may not be effectively utilizing additional contextual information during training.  

The largest improvements arise when retrieval is combined with multi-candidate selection. The reranking stage improves pass@1 over the best non-reranked method by approximately 4 points on HumanEval and 12 points on MBPP across open models, suggesting that retrieval should be treated as a mechanism for generating diverse candidate solutions, after which selection becomes the primary bottleneck. The oracle (``Ideal Reranker'') indicates additional headroom, especially on HumanEval ($\approx$10-point average gap), implying that the candidate sets frequently include correct solutions that are not selected by the current reranker.

\begin{framed}
\textbf{Answer to RQ1:} Code-centric PKG improves code generation when retrieval units are code-structured and when candidate selection is used; naïve retrieval can harm performance. Smaller retrieval units yield the best average performance.
\end{framed}

\textbf{RQ2: Does text-centric PKG improve code generation?}
In this research question, we investigate the potential of leveraging text-centric data to improve code generation. To be more specific, we built a PKG on the Tutorials dataset \citep{wang2024coderag}, which was processed using the Gemma2-9B model \cite{team2024gemma}, producing hierarchical JSON representations, enabling us to generate a PKG as explained in Section \ref{sec:pkg-generation}. Our baselines include: 1) None: No retrieval-augmented generation, 2) BM25: Applied to the entire dataset without pre-processing, 3) VoyageEmb: Embeddings from question-answer pairs for retrieval, and 4) JSON-PKG: Our proposed PKG based on text-centric data.

Tables~\ref{table:pkg_json_human-eval} and~\ref{table:pkg_json_mbpp} show the evaluation results. Relative to no retrieval, JSON-PKG yields heterogeneous effects on HumanEval. It improves pass@1 for CodeLlama-7B (+3 points), Llama3.1-8B (+8 points), and StarCoder2-7B (+16 points), but degrades performance for CodeLlama-13B (-1 point) and DeepSeek-Coder-7B (-5 points). On MBPP, JSON-PKG yields consistent but modest improvements across all evaluated models, ranging from +1 to +8 points, with the largest gain for Llama3.1-8B (+8 points).

Baselines that retrieve from unstructured tutorial rows remain unstable. BM25 reduces the performance on HumanEval and slightly on MBPP, while dense retrieval (VoyageEmb) improves the performance on HumanEval slightly but on average reduces the performance of the models on MBPP. Overall, these results indicate that retrieval effectiveness depends not only on the retriever, but also on the structure and relevance of the indexed artifacts.

Comparing JSON-PKG (text-centric) against Block-PKG (code-centric) from Tables~\ref{table:different_methods_human-eval} and~\ref{table:different_methods_mbpp}, text-centric retrieval is, on average, slightly weaker on HumanEval but competitive on MBPP. On HumanEval, JSON-PKG (pass@1=53.2\%) trails Block-PKG (pass@1=55.0\%) by 1.8 points on average, but it outperforms Block-PKG for Llama3.1-8B and StarCoder2-7B. On MBPP, JSON-PKG is approximately 0.6 points higher on average, largely because Block-PKG substantially degrades DeepSeek-Coder-7B on MBPP while JSON-PKG yields a small improvement for the same model. This suggests that tutorial-based retrieval can be beneficial when code-based retrieval surfaces misleading context.

The results support a precision-relevance tradeoff for text-centric retrieval. Tutorial content can provide high-level scaffolding such as algorithm outlines, invariants, and edge-case reasoning that improves generation, which is consistent with the large gains observed for StarCoder2-7B and Llama3.1-8B on HumanEval. However, unstructured tutorial text can also introduce noise. Retrieved explanations may be topically related but not operationally aligned with the benchmark specification, and this can bias the model toward an incorrect implementation pattern or away from benchmark-specific constraints. This is consistent with the degradations observed for CodeLlama-13B and DeepSeek-Coder-7B on HumanEval.

Taken together, the evidence indicates that text-centric PKG retrieval can improve code generation, but its benefits are model- and benchmark-dependent. We hypothesize that general LLMs benefit more from text-centric data as supplementary context than CLLMs. Additionally, while we tested retrieval methods like BM25 and VoyageEmb, JSON-PKG outperformed them in Pass@1 accuracy across both benchmarks. However, when comparing Block-PKG with JSON-PKG, code-centric data still offers greater benefits for code generation tasks, highlighting that code-focused data remains more effective for these specific tasks.

\begin{framed}
    \textbf{Answer to RQ2.} Text-centric PKG can improve code generation, but the effect is not uniform across models. While general LLMs benefit more from text-centric data, Code-LLMs benefit more from code-centric data.
\end{framed}

\begin{table}[h!]
\centering
\caption{The performance of PKG on HumanEval, using tutorials data, is reported as pass@1. Red cells indicate accuracy below NoRAG, green cells above, with color intensity reflecting significance.}
\vspace{2pt}
\renewcommand{\arraystretch}{1.3}
\scalebox{0.70}{
\begin{tabular}{lcccccccc}
\toprule
\textbf{Model} & \textbf{None} & \textbf{BM25} & \textbf{VoyageEmb} & \textbf{JSON-PKG}  \\ 
\midrule
CodeLlama-7B      & 33\% \cellcolor{gray!0} & \cellcolor{OrangeRed!10} 28\% &\cellcolor{ForestGreen!5}35\%& \cellcolor{ForestGreen!6} 36\%  \\ 
CodeLlama-13B     & 42\% \cellcolor{gray!0} & \cellcolor{OrangeRed!26}29\% &\cellcolor{ForestGreen!3}43\%& \cellcolor{OrangeRed!2}41\% \\ 
Llama3.1-8B       & 55\% \cellcolor{gray!0} & \cellcolor{OrangeRed!16}47\% &\cellcolor{ForestGreen!8}58\%& \cellcolor{ForestGreen!12}63\% \\ 
StarCoder2-7B     & 45\% \cellcolor{gray!0} & \cellcolor{OrangeRed!10}40\% &\cellcolor{ForestGreen!25}59\%& \cellcolor{ForestGreen!30}61\% \\ 
DeepSeek-Coder-7B & 70\% \cellcolor{gray!0} &  \cellcolor{OrangeRed!20}60\% &\cellcolor{OrangeRed!20}59\%& \cellcolor{OrangeRed!10}65\%  \\ 
\textbf{Avg.}     & \textbf{49.0\%} & \textbf{40.8\%} & \textbf{50.8\%} & \textbf{53.2\%} \\
\bottomrule
\end{tabular}}
\label{table:pkg_json_human-eval}
\end{table}

\begin{table}[h!]
\centering
\caption{PKG performance on MBPP using tutorial data, measured by pass@1. Red indicates accuracy below NoRAG, green above, with shading intensity showing significance.}
\vspace{2pt}
\renewcommand{\arraystretch}{1.3}
\scalebox{0.70}{
\begin{tabular}{lcccccccc}
\toprule
\textbf{Model} & \textbf{None} & \textbf{BM25} & \textbf{VoyageEmb} & \textbf{JSON-PKG}  \\ 
\midrule
CodeLlama-7B      & 38\% \cellcolor{gray!0} & \cellcolor{OrangeRed!10} 29\% &\cellcolor{OrangeRed!10}30\%& \cellcolor{ForestGreen!6} 41\%  \\ 
CodeLlama-13B     & 44\% \cellcolor{gray!0} & \cellcolor{OrangeRed!20}35\% &\cellcolor{OrangeRed!20}36\%& \cellcolor{ForestGreen!2}45\% \\ 
Llama3.1-8B       & 43\% \cellcolor{gray!0} & \cellcolor{ForestGreen!12}48\% &\cellcolor{ForestGreen!20}51\%& \cellcolor{ForestGreen!20}51\% \\ 
StarCoder2-7B     & 46\% \cellcolor{gray!0} & \cellcolor{ForestGreen!5}49\% &\cellcolor{OrangeRed!16}38\%& \cellcolor{ForestGreen!12}50\% \\ 
DeepSeek-Coder-7B & 56\% \cellcolor{ForestGreen!0} &  \cellcolor{ForestGreen!7}58\% &\cellcolor{OrangeRed!16}51\%& \cellcolor{ForestGreen!5}57\%  \\ 
\textbf{Avg.}     & \textbf{45.4\%} & \textbf{43.8\%} & \textbf{41.2\%} & \textbf{48.8\%} \\
\bottomrule
\end{tabular}}
\label{table:pkg_json_mbpp}
\end{table}

\textbf{RQ3: Which knowledge representation method is most effective in optimizing context retrieval for code generation tasks?}

To isolate the impact of representation, we compare three families of retrieval units: (i) row/Q\&A representations, (ii) function-level representations, and (iii) block-level representations, as shown in Tables \ref{table:different_methods_human-eval} and \ref{table:different_methods_mbpp}. Across open models, row/Q\&A representations are the least reliable: both sparse and dense retrievers over these units often underperform no retrieval, indicating that ``retrieval quality” is not just a function of the retriever algorithm but also of the granularity and cleanliness of the indexed artifacts.

Function-level representations mitigate some of this instability but do not dominate uniformly, implying that simply extracting functions is insufficient when retrieved functions are long or only partially relevant. Block-level representations generally provide stronger non-reranked performance and larger gains over function-level retrieval on MBPP, but can still fail for certain models. This failure mode is consistent with a precision–recall tradeoff: higher granularity increases the probability of retrieving highly relevant fragments, but also increases exposure to superficially similar fragments that trigger incorrect solution templates.

Finally, reranking consistently dominates representation-only improvements, indicating that the most effective system is not “the best single context,” but rather a pipeline that uses representation to generate candidate diversity and uses selection to control retrieval-induced error.

\begin{framed}
    \textbf{Answer to RQ3:} Code-structured representations (functions/blocks) are generally more effective than row/Q\&A, but the decisive factor in these results is robust selection (reranking), not representation alone.
\end{framed}

\begin{figure}[htbp]
    \centering
    \includegraphics[width=1\columnwidth]{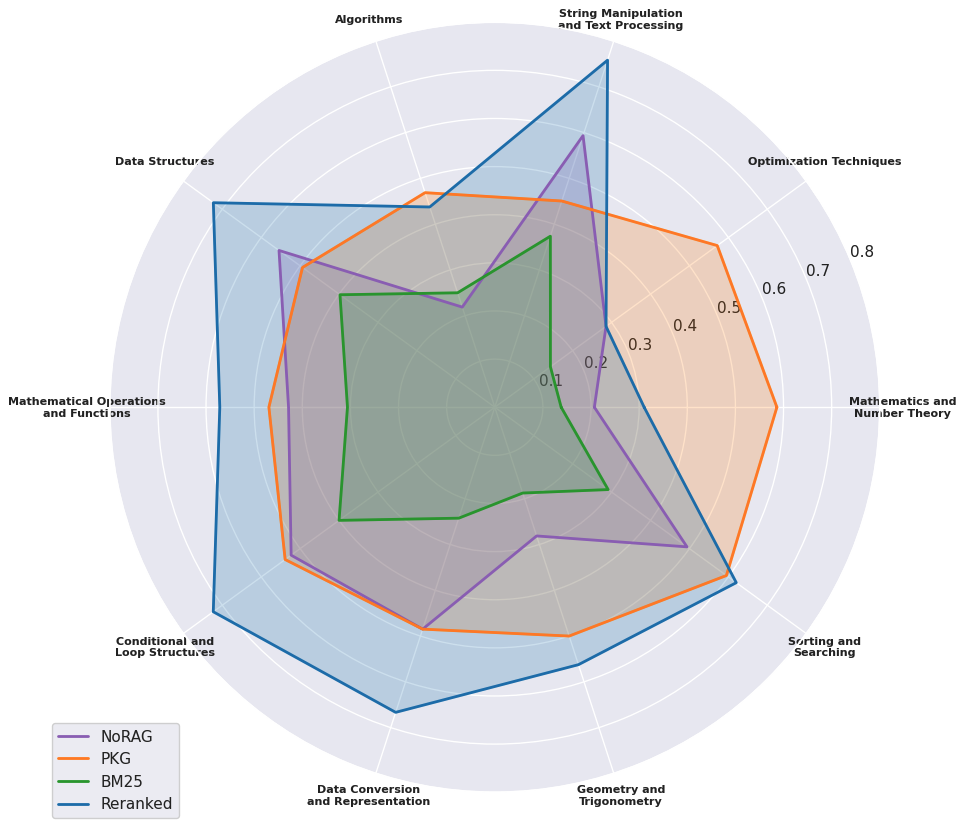}
    \caption{Comparison of different approaches across 10 topics using the MBPP benchmark on StarCoder2-7B }
    \Description{Comparison of different approaches across 10 topics using the MBPP benchmark on StarCoder2-7B }
    \label{fig:topic_based_accuracy}
\end{figure}

\section{Discussions}
\label{sec:discussions}

This section synthesizes the empirical findings into actionable insights. We first explore the impact of PKG on closed-source models (Section \ref{sec:disc-closed-source-models}). Rather than focusing only on aggregate pass@1, we examine how retrieval changes behavior across problem topics (Section \ref{sec:disc-topics}) and error modes (Section \ref{sec:disc-errors}), and what these changes imply for designing reliable retrieval-augmented code generation systems. We then contextualize accuracy gains with preprocessing and storage costs (Section \ref{sec:disc-cost}). Section \ref{sec:disc-implications} discusses implications for researchers and practitioners regarding when PKG-based retrieval is beneficial, when it can be harmful, and what system components most strongly determine robustness. Finally, Section \ref{sec:disc-gaps} discusses open gaps and limitations.


\subsection{Extending to Closed-source Models}
\label{sec:disc-closed-source-models}

We explore the effect of PKG on closed-source models. As shown in Tables \ref{table:different_methods_human-eval_prop} and \ref{table:different_methods_mbpp_prop}, for commercial models, the marginal benefit of retrieval is substantially smaller and sometimes negative. Given their high no-retrieval baselines (e.g., 96.3 pass@1 on HumanEval for Claude Sonnet 4), retrieval likely provides limited novel information while still incurring risks from irrelevant or contradictory context. This suggests that retrieval policies may need to be model-adaptive, with conservative gating for powerful, closed-source models and more diverse candidates for smaller, open-source models.
Note that still even closed-source models can benefit from the Func-PKG and reranker proposed by our work. The most significant improvements are seen for GPT-4o and GPT-4o-mini, where the results are improved by 2 and 2.8 pass@1 scores, respectively, when the reranking is applied. 

\begin{table*}[!ht]
\centering
\caption{Performance of closed-source models with code-centric retrieval-augmented code generation on HumanEval, reported as pass@1. Red cells indicate performance below NoRAG and green cells show scores above NoRAG, with color intensity reflecting significance. ``Ideal Reranker'' serves as an oracle upper bound for our proposed re-ranking stage.}
\vspace{2pt}
\renewcommand{\arraystretch}{1.3}
\scalebox{0.70}{
\begin{tabular}{l l cccccccc}
\toprule
\textbf{} & \textbf{Model} & \textbf{None} & \textbf{BM25} & \textbf{VoyageEmb} & \textbf{Func-BM25} & \textbf{Func-PKG} & \textbf{Block-PKG} & \textbf{Reranked} & \textbf{Ideal Reranker} \\ 
\midrule

& Claude-3-Haiku & \cellcolor{gray!0}\textbf{74.4\%} & \cellcolor{OrangeRed!37}72.0\% & \cellcolor{OrangeRed!9}73.8\% & N/A & \cellcolor{OrangeRed!46}71.3\% & \cellcolor{OrangeRed!90}67.7\% & \cellcolor{OrangeRed!9}73.8\% & \cellcolor{BlueViolet!20}87.8\% \\
& Claude-Sonnet-4 & \cellcolor{gray!0}\textbf{96.3\%} & \cellcolor{OrangeRed!18}95.1\% & \cellcolor{OrangeRed!18}95.1\% & N/A & \cellcolor{OrangeRed!18}95.1\% & \cellcolor{OrangeRed!27}94.5\% & \cellcolor{OrangeRed!9}95.7\% & \cellcolor{BlueViolet!20}100.0\% \\
& GPT-4o & 88.4\% \cellcolor{gray!0} & \cellcolor{ForestGreen!9}89.0\% & \cellcolor{ForestGreen!18}\textbf{89.6\%} & N/A & \cellcolor{ForestGreen!9}89.0\% & \cellcolor{ForestGreen!18}89.6\% & \cellcolor{ForestGreen!18}89.6\% & \cellcolor{BlueViolet!20}95.7\% \\
& GPT-4o-mini & 87.2\% \cellcolor{gray!0} & \cellcolor{OrangeRed!64}82.9\% & \cellcolor{OrangeRed!27}85.4\% & N/A & \cellcolor{OrangeRed!18}86.0\% & \cellcolor{OrangeRed!55}83.5\% & \cellcolor{ForestGreen!9}\textbf{87.8\%} & \cellcolor{BlueViolet!20}95.1\% \\
& \textbf{Avg.} & \textbf{86.6\%} & \textbf{84.8\%} & \textbf{86.0\%} & \textbf{N/A} & \textbf{85.4\%} & \textbf{83.8\%} & \textbf{86.7\%} & \textbf{94.7\%} \\

\bottomrule
\end{tabular}}
\label{table:different_methods_human-eval_prop}
\end{table*}

\begin{table*}[t!]
\centering
\caption{Performance of closed-source models with code-centric retrieval-augmented code generation on MBPP, reported as pass@1. Red cells indicate accuracy below NoRAG, green cells indicate accuracy above, and color intensity reflects significance. ``Ideal Re-ranker'' serves as the upper bound for the proposed re-ranker method.}
\label{table:different_methods_mbpp_prop}
\vspace{6pt}
\renewcommand{\arraystretch}{1.3}
\scalebox{0.70}{
\begin{tabular}{l l cccccccc}
\toprule
\textbf{} & \textbf{Model} & \textbf{None} & \textbf{BM25} & \textbf{VoyageEmb} & \textbf{Func-BM25} & \textbf{Func-PKG} & \textbf{Block-PKG} & \textbf{Reranked} & \textbf{Ideal Reranker} \\ 
\midrule

& Claude-3-Haiku & 67.2\% \cellcolor{gray!0} & \cellcolor{ForestGreen!6}\textbf{67.6\%} & \cellcolor{OrangeRed!18}66.0\% & N/A & \cellcolor{ForestGreen!6}67.6\% & \cellcolor{OrangeRed!3}67.0\% & \cellcolor{ForestGreen!6}67.6\% & \cellcolor{BlueViolet!20}73.0\% \\
& Claude-Sonnet-4 & \cellcolor{gray!0}\textbf{83.6\%} & \cellcolor{OrangeRed!18}82.4\% & \cellcolor{OrangeRed!21}82.2\% & N/A & \cellcolor{OrangeRed!6}83.2\% & \cellcolor{OrangeRed!9}83.0\% & \cellcolor{OrangeRed!21}82.2\% & \cellcolor{BlueViolet!20}87.4\% \\
& GPT-4o & 81.4\% \cellcolor{gray!0} & \cellcolor{OrangeRed!12}80.6\% & \cellcolor{ForestGreen!24}83.0\% & N/A & \cellcolor{ForestGreen!3}81.6\% & 81.4\% & \cellcolor{ForestGreen!30}\textbf{83.4\%} & \cellcolor{BlueViolet!20}86.0\% \\
& GPT-4o-mini & 74.2\% \cellcolor{gray!0} & \cellcolor{OrangeRed!3}74.0\% & \cellcolor{OrangeRed!12}73.4\% & N/A & \cellcolor{ForestGreen!6}74.6\% & \cellcolor{OrangeRed!6}73.8\% & \cellcolor{ForestGreen!45}\textbf{77.2\%} & \cellcolor{BlueViolet!20}79.2\% \\
& \textbf{Avg.} & \textbf{76.6\%} & \textbf{76.2\%} & \textbf{76.2\%} & \textbf{N/A} & \textbf{76.8\%} & \textbf{76.3\%} & \textbf{77.6\%} & \textbf{81.4\%} \\

\bottomrule
\end{tabular}
}
\end{table*}

\subsection{Problem Topics Benefiting from RAG}
\label{sec:disc-topics}

We analyze topic-level performance on MBPP using DeepSeek-Coder-7B. We map 134 MBPP unique categories into 10 broader topics and report pass@1 per topic for four settings: NoRAG, BM25, PKG, and Re-ranked.
Figure~\ref{fig:topic_based_accuracy} shows three consistent patterns.
First, PKG outperforms BM25 across all topics. The improvement is largest in topics where lexical matching is least reliable, including \textit{Mathematics and Number Theory} and \textit{Optimization Techniques}, where BM25 is low while PKG is substantially higher. This indicates that topic-relevant retrieval requires more than surface-form overlap, and that PKG retrieves context that better matches the semantic intent of the tasks.

Second, relative to NoRAG, PKG improves pass@1 in 7 out of 10 topics. Gains are most visible for topics such as \textit{Mathematics and Number Theory}, \textit{Optimization Techniques}, \textit{Geometry and Trigonometry}, and \textit{Sorting and Searching}. In contrast, PKG underperforms NoRAG for \textit{String Manipulation and Text Processing} and \textit{Data Structures}. For these two topics, results indicate that retrieved context can be net distracting even when it is broadly relevant at the topic level.

Third, re-ranking changes the topic profile. Re-ranked achieves the highest pass@1 in most topics, with prominent gains in \textit{String Manipulation and Text Processing}, \textit{Data Structures}, \textit{Conditional and Loop Structures}, and \textit{Data Conversion and Representation}. However, re-ranking is not uniformly beneficial. In \textit{Mathematics and Number Theory}, \textit{Optimization Techniques}, and \textit{Algorithms}, the Re-ranked curve falls below PKG, suggesting that the re-ranker can select suboptimal candidates in these domains even when retrieval produces useful options.

The topic-level differences suggest that retrieval provides value through two mechanisms, and that their relative importance varies by topic.
In topics such as sorting, geometry, and control-flow patterns, correct solutions often reuse recognizable program schemas. Retrieval can supply concrete implementation details (for example, standard loops, boundary handling, or library usage), and re-ranking can exploit candidate diversity to select a correct instantiation. This is consistent with the large separation between Re-ranked and NoRAG in these topics.

In contrast, string processing and data-structure tasks are often sensitive to small semantic details, including off-by-one behavior, formatting constraints, and invariants that are easy to violate by adapting a superficially similar snippet. If retrieved context encourages a near-match template, the model can become less faithful to the exact specification, which is consistent with PKG underperforming NoRAG in these two topics. The fact that re-ranking strongly improves these topics suggests that retrieval still produces some correct candidates, but selection becomes the bottleneck, the system benefits when it can choose among multiple solutions and avoid the misleading template.

The re-ranker underperformance in mathematics, optimization, and algorithms relative to PKG indicates a different failure mode. A plausible explanation is that candidate sets in these topics contain many solutions that are syntactically well-formed and structurally plausible but mathematically incorrect, and the re-ranker scores correlate with surface plausibility rather than semantic validity. This motivates adding topic-aware signals, such as lightweight symbolic checks, constraint extraction, or re-ranker features that better reflect semantic correctness for mathematically constrained tasks.

\begin{figure}[htbp]
    \centering
    \includegraphics[width=1.0\textwidth]{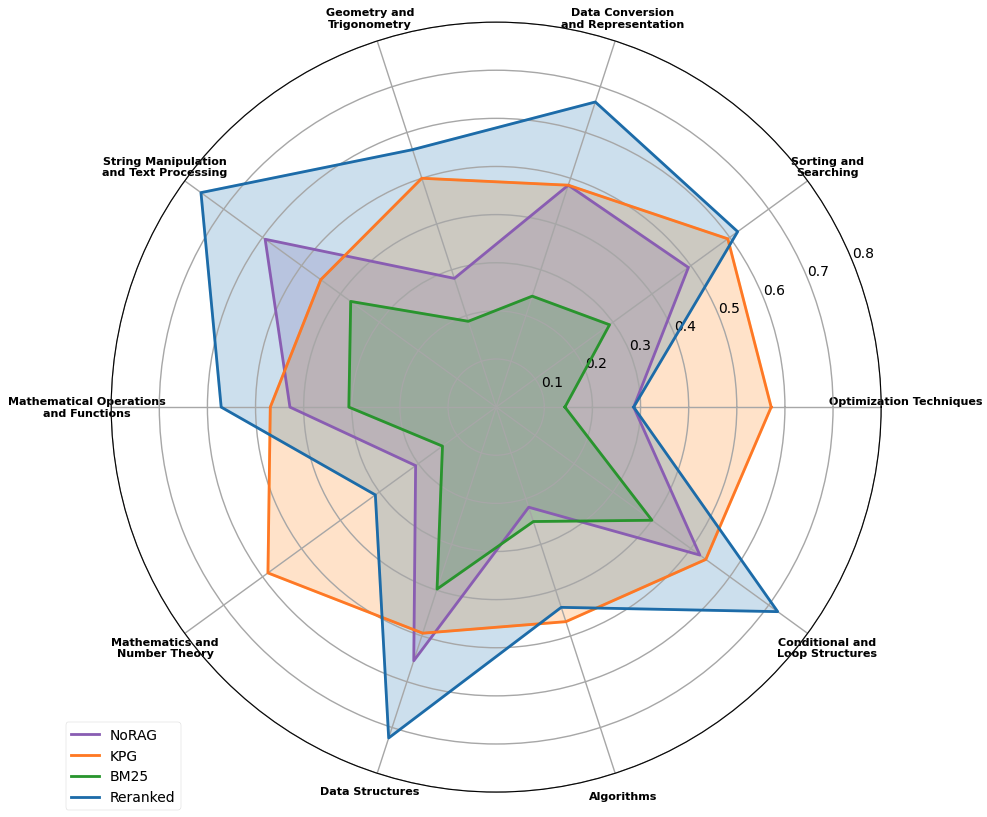}
    \caption{Comparison of different approaches across 10 topics using the MBPP benchmark on CodeLlama-7B. }
    \Description{Comparison of different approaches across 10 topics using the MBPP benchmark on CodeLlama-7B. }
    \label{fig:topic_based_accuracy_codellama}
\end{figure}

Figure \ref{fig:topic_based_accuracy_codellama} presents the pass@1 accuracy for each method—NoRAG, PKG, BM25, and the re-ranked approach—across various programming topics. Similar to the performance observed with the StarCoder2-7B model, the re-ranker struggles to correctly prioritize solutions in the 'Optimization Techniques,' 'Mathematics,' and 'Algorithm' categories. However, in other topic areas, the re-ranker demonstrates superior performance compared to the other methods. Notably, for this model, PKG achieves higher accuracy across most topics, with the exception of 'String Manipulation' and 'Data Structures,' where it is outperformed by other approaches.

\begin{figure}[htbp]
    \centering
    \includegraphics[width=1.0\textwidth]{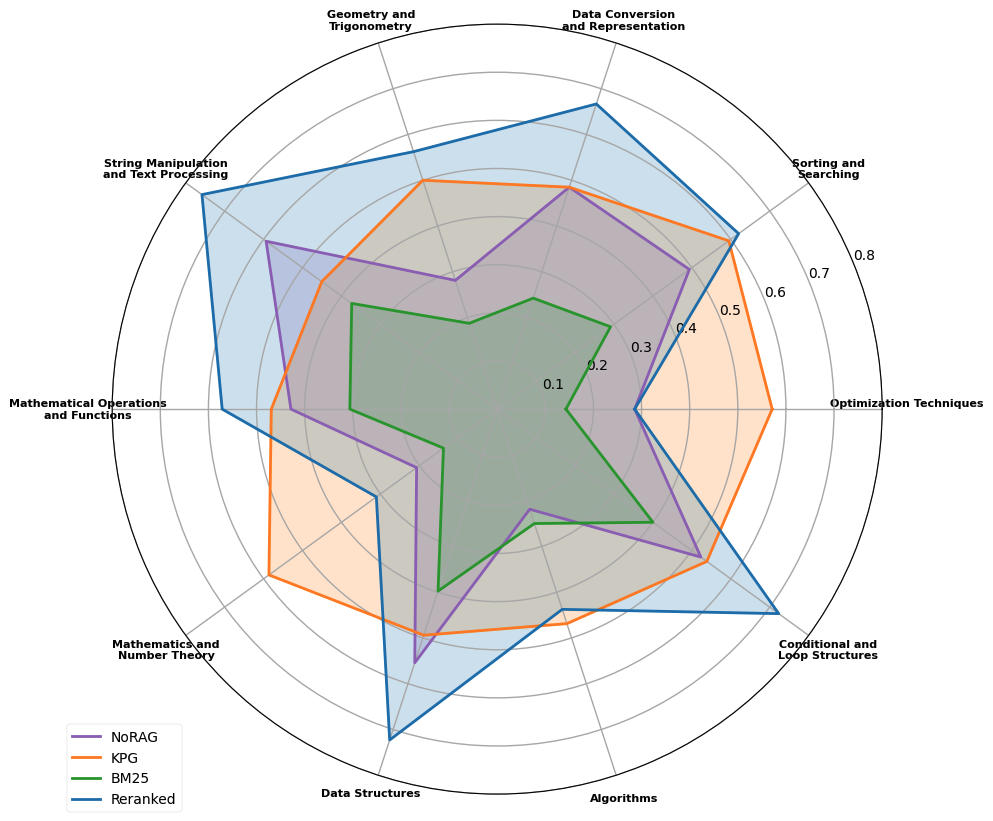}
    \Description{Comparison of different approaches across 10 topics using the MBPP benchmark on DeepSeek-Coder-7B }
    \caption{Comparison of different approaches across 10 topics using the MBPP benchmark on DeepSeek-Coder-7B }
    \label{fig:topic_based_accuracy_deepseek}
\end{figure}

Figure \ref{fig:topic_based_accuracy_deepseek} illustrates the Pass@1 accuracy for each evaluation method: NoRAG, PKG, BM25, and the re-ranked approach, across a range of programming topics. The performance trends observed with the DeepSeek-Coder-7B model are echoed here. Specifically, the re-ranking method shows difficulty in accurately prioritizing solutions within the categories of 'Optimization Techniques,' 'Mathematics,' and 'Algorithms.' Despite these challenges, the re-ranked approach excels in other topic areas, demonstrating superior performance compared to the other methods.

Notably, the PKG method achieves higher accuracy across most topics evaluated. However, it does face competition in the 'String Manipulation' and 'Data Structures' categories, where it is outperformed by NoRAG approach. We have observed the same behaviour for the previous models.

In summary, on MBPP, PKG improves performance over BM25 for all topics and improves over NoRAG for most topics, but it is not uniformly beneficial. Topics that rely on reusable program schemas benefit strongly from retrieval and re-ranking, while topics that are sensitive to fine-grained semantic constraints can experience retrieval-induced regressions unless the system includes robust candidate selection.

\subsection{Error Types Reduced or Introduced by Applying RAG}
\label{sec:disc-errors}

Table~\ref{table:error_analysis} summarizes exception types observed on MBPP for three models with and without Block-PKG. MBPP is chosen as it is a more complex dataset compared to HumanEval. The raw counts show that retrieval changes not only whether a solution is correct, but also how it fails when it is incorrect.
For StarCoder-7B, Block-PKG reduces AssertionErrors from 198 to 147 ($-51$) and reduces SyntaxErrors from 2 to 0 ($-2$), but increases NameErrors from 51 to 64 ($+13$) and introduces IndentationErrors from 0 to 18 ($+18$). For CodeLlama-7B, Block-PKG reduces NameErrors from 138 to 65 ($-73$) and reduces AssertionErrors from 180 to 162 ($-18$), but increases TypeErrors from 28 to 37 ($+9$) and introduces SyntaxErrors from 0 to 1 ($+1$). For DeepSeekCoder-7B, Block-PKG increases AssertionErrors from 135 to 146 ($+11$), increases NameErrors from 64 to 78 ($+14$), and increases TypeErrors from 4 to 16 ($+12$).
The total number of recorded errors (summed across listed exception types) decreases for StarCoder-7B (265 to 244, $-21$) and CodeLlama-7B (357 to 269, $-88$), but increases for DeepSeekCoder-7B (208 to 249, $+41$).

\begin{table*}[t!]
\centering
\caption{Error Analysis on MBPP for different settings}
\scalebox{0.7}{
\begin{tabular}{@{}l *{6}{c}@{}} 
\toprule
Error Type 
    & StarCoder-7B 
    & \begin{tabular}{@{}c@{}} StarCoder-7B \\ + Block-PKG \end{tabular}  
    & CodeLlama-7B 
    & \begin{tabular}{@{}c@{}} CodeLlama-7B \\ + Block-PKG \end{tabular} 
    & DeepSeekCoder-7B 
    & \begin{tabular}{@{}c@{}} DeepSeekCoder-7B \\ + Block-PKG \end{tabular} \\ 
\midrule
\# of AssertionErrors  & 198 & 147 & 180 & 162 & 135 & 146 \\ 
\# of NameErrors       &  51 &  64 & 138 &  65 &  64 &  78 \\ 
\# of TypeErrors       &  11 &   8 &  28 &  37 &   4 &  16 \\ 
\# of SyntaxErrors     &   2 &   0 &   0 &   1 &   0 &   0 \\ 
\# of IndentationErrors &   0 &  18 &   0 &   0 &   0 &   0 \\ 
\# of Others           &   3 &   7 &  11 &   4 &   5 &   9 \\ 
\bottomrule
\end{tabular}
}
\label{table:error_analysis}
\end{table*}

Two consistent mechanisms explain the observed shifts.
First, retrieval can reduce semantic mismatch failures, which often manifest as AssertionErrors and NameErrors. When retrieved context provides relevant function signatures, variable naming conventions, or common implementation patterns, models may produce code that better matches the target specification and avoids undefined identifiers. This is consistent with the large NameError reduction for CodeLlama-7B and the AssertionError reduction for StarCoder-7B.

Second, retrieval can introduce formatting and compatibility failures. The appearance of IndentationErrors for StarCoder-7B suggests that injected code blocks, or the model's copying of retrieved code, can disrupt indentation structure. The increase in TypeErrors for CodeLlama-7B and DeepSeekCoder-7B is consistent with models adopting retrieved templates that assume different input types or data conventions than the benchmark requires.

The DeepSeekCoder-7B pattern is qualitatively different: error counts increase across major categories. This should not be attributed to an inability to use context without additional evidence. A simpler explanation is that retrieval reduces correctness for this model under Block-PKG (as also suggested by the aggregate MBPP results), thereby increasing the number of failing executions and the opportunity for errors to occur.

On MBPP, retrieval can reduce failures consistent with semantic mismatch (notably AssertionErrors and NameErrors for some models), but it can introduce new failure modes, particularly formatting errors (IndentationErrors) and type mismatches (TypeErrors). The direction and magnitude of these effects are model-dependent.

\begin{figure}[htbp]
    \centering
    \includegraphics[width=1\columnwidth]{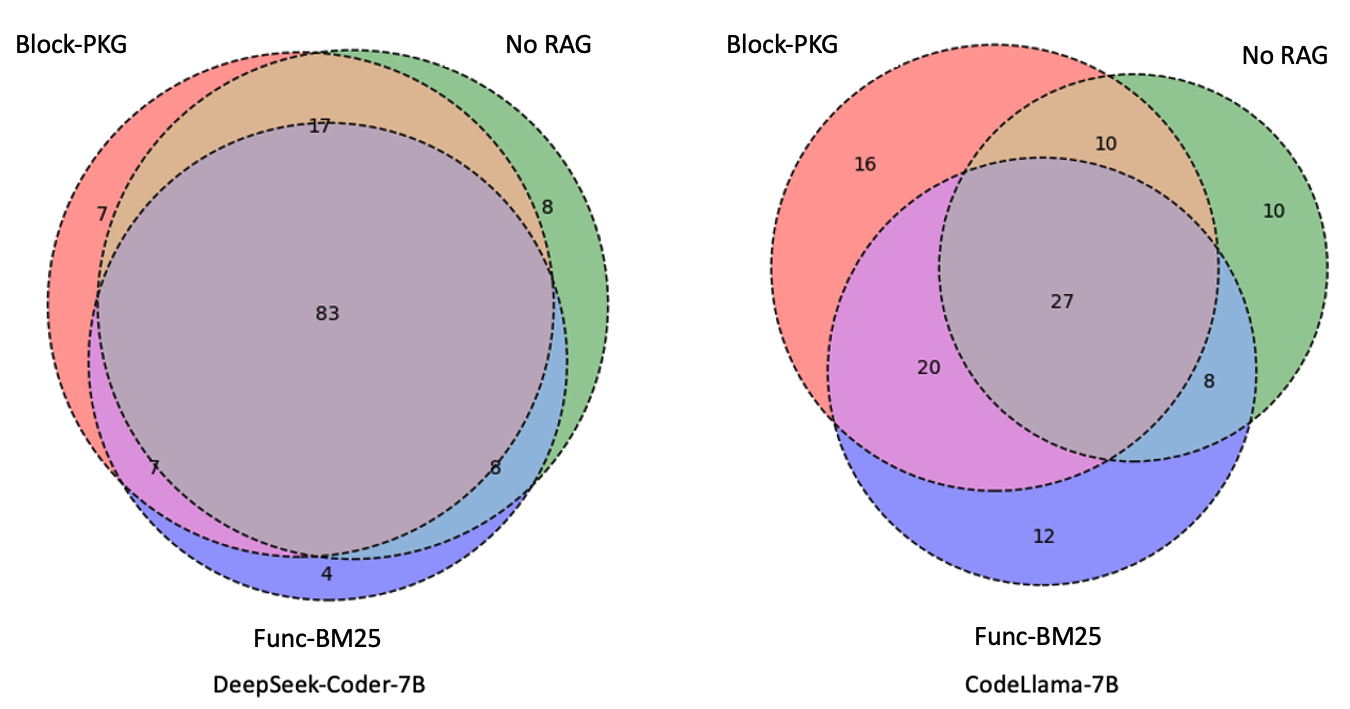}
    \caption{This figure illustrates the impact of three approaches -- our technique, Programming Knowledge Graph (Block-PKG), Func-BM25, and NoRAG -- on solving HumanEval problems using the DeepSeek-Coder-7B (left) and CodeLlama-7B (right) models. Considering CodeLlama-7B, it shows that 16 problems were uniquely solved by the PKG, 12 problems by Func-BM25, and 27 problems were solved by all three approaches. }
    \Description{This figure illustrates the impact of three approaches -- our technique, Programming Knowledge Graph (Block-PKG), Func-BM25, and NoRAG -- on solving HumanEval problems using the DeepSeek-Coder-7B (left) and CodeLlama-7B (right) models. Considering CodeLlama-7B, it shows that 16 problems were uniquely solved by the PKG, 12 problems by Func-BM25, and 27 problems were solved by all three approaches. }
    \label{fig:venn_comp}
\end{figure}

As shown in Figure \ref{fig:venn_comp}, different approaches excel at solving distinct types of problems, demonstrating the need for a re-ranker. When the initial retrieved content introduces hallucinations into the output, the re-ranker can prioritize solutions generated without relying on RAG-based content, reducing the influence of erroneous data.

\subsection{Cost Trade-off}
\label{sec:disc-cost}


We compare PKG’s generation cost with a standard RAG setup using VoyageAI and BM25 retrieval on the PythonAlpaca dataset \cite{nic2024pyalpaca} in Table \ref{table:comparison}. PKG and VoyageAI share identical encoding times since both use the same embedding model (Voyage-Code2) and dataset. Unlike embedding-based RAG methods, PKG requires an additional hour of processing but achieves a 9.4\% higher accuracy on average. Neo4j’s semantic vector indexing enables efficient graph updates with logarithmic complexity: $O(\log N)$ for nodes and $O(\log M)$ for relationships. Retrieval involves comparing query embeddings with all nodes, resulting in $O(N \cdot d)$ complexity, where $d$ is the embedding dimension. In practice, queries took about 3 seconds each.

\begin{table}[h!]
\centering
\caption{Time and storage usage for creating RAG data sources on PythonAlpaca \cite{nic2024pyalpaca}. Time is in minutes, and storage (last row) is in megabytes (MB).}
\label{table:comparison}
\scalebox{0.85}{
\begin{tabular}{@{}lccc@{}}
\toprule
\textbf{Step} & \textbf{PKG} & \textbf{VoyageAI} & \textbf{BM25} \\ \midrule
Python Code Extraction &  3 & - & - \\
Block Extraction        &  25 & - & - \\
Encoding                &  241 & 240 & 44 \\
Neo4j Graph Generation  &  33 & - & - \\ \midrule
\textbf{Overall Time}    & \textbf{301} & \textbf{241} & \textbf{44} \\
\textbf{Storage Usage (MB)}  & \textbf{12,530} & \textbf{8,440} & \textbf{315} \\ \bottomrule
\end{tabular}
}
\end{table}

Although PKG requires more time for processing, computational inference costs is reduced, while performance is increased. 
Table \ref{table:token_budget} presents the average token length of the additional context across different retrieval settings. As shown in the table, Block-PKG has the lowest number of tokens compared to others. 
One of the key advantages of retrieving finer-grained contextual information is the ability to provide the model with a reduced token budget, thereby lowering computational inference costs and minimizing monetary expenses, particularly for proprietary models. 

\begin{table*}[h!]
\centering
\caption{The average number of additional tokens in different approaches for the HumanEval benchmark.}
\vspace{2pt}
\renewcommand{\arraystretch}{1.3}
\scalebox{0.70}{
\begin{tabular}{lcc}
\toprule
\textbf{RAG Method} & \textbf{Avg. Tokens (CodeLlama)} & \textbf{Avg. Tokens (DeepSeek)} \\ 
\midrule
\textbf{Block-PKG}      & \textbf{87} & \textbf{84} \\ 
Func-PKG     & 188 & 182 \\ 
BM-25       & 226 & 218 \\ 
Voyage     & 349 & 339 \\ 
\bottomrule
\end{tabular}}
\label{table:token_budget}
\end{table*}

\subsection{Implications for Researchers and Practitioners}
\label{sec:disc-implications}
    
Our results suggest that retrieval-augmented code generation is best viewed as a coupled system with three interacting design choices: (i) the \emph{unit of retrieval} (rows versus functions versus blocks versus path-values), (ii) the \emph{selection mechanism} (single-shot augmentation versus candidate selection), and (iii) the \emph{budgeting policy} (how much retrieval and how many candidates are permitted). Below we summarize implications that follow directly from our findings and analysis.

\paragraph{Implication 1: Treat retrieval as a high-variance intervention that requires safeguards.}
Across models and topics, retrieval can both improve and degrade outcomes. Topic-level results show that even a semantically motivated retriever (PKG) can underperform NoRAG in categories that are sensitive to specification details, such as string processing and data-structure tasks. Error analysis further shows that retrieval can introduce new failure modes, including indentation and type errors, that are absent or rare without retrieval. For practitioners, this implies that retrieval should not be enabled unconditionally. A practical deployment strategy is to use retrieval gating based on simple signals, for example, predicted relevance, topic classifier confidence, or a heuristic that detects when retrieved code is likely to conflict with the query constraints, and to fall back to NoRAG when uncertainty is high.

\paragraph{Implication 2: Granularity is not a cosmetic choice; it changes the failure modes.}
Coarser units provide broader context but increase the risk of irrelevant content, while finer units can increase precision but also increase exposure to superficially similar templates. The topic analysis suggests that tasks with reusable schemas benefit from fine-grained retrieval and candidate selection, whereas tasks with tight semantic constraints can be harmed by near-miss templates. For researchers, this implies that reporting only overall pass@1 can hide important differences. Studies should report performance stratified by topic or difficulty and should include a help-hurt analysis to characterize regressions induced by retrieval at different granularities.

\paragraph{Implication 3: Selection is a first-class research problem, not an implementation detail.}
Candidate selection changes both accuracy and error types. The topic-level results show that reranking can substantially improve categories that otherwise regress under retrieval, indicating that many failures are selection failures rather than retrieval failures. At the same time, reranking underperforms PKG in topics such as mathematics and algorithms, suggesting that similarity-based selection can correlate with surface similarities rather than semantic validity. This motivates reranking objectives that are closer to correctness, including execution-aware signals, constraint checks, or rerankers trained on counterexamples where plausible solutions are incorrect. For practitioners, this implies that single-prompt augmentation is unlikely to be robust without some form of candidate selection.

\paragraph{Implication 4: Error-mode shifts can guide targeted mitigation.}
Execution-trace errors provide actionable feedback about what retrieval changes. For example, indentation errors point to formatting issues in how retrieved code is inserted into the prompt, and type errors suggest mismatches between retrieved templates and the task specification. These signals can be used to design defensive post-processing and formatting rules, such as canonical indentation normalization, strict delimiter templates, and lightweight static checks before execution. For researchers, this implies that evaluation should report not only accuracy but also how retrieval redistributes failure types, since two systems with similar pass@1 may have very different reliability characteristics.

\paragraph{Implication 5: Indexing and preprocessing costs are acceptable when amortized, but must be justified by downstream gains.}
The cost analysis shows that PKG construction introduces additional preprocessing steps beyond embedding-based retrieval, increasing both build time and storage footprint. However, these costs are amortized across many queries and can be justified when the downstream accuracy improvements are large enough and when updates are incremental. For practitioners, this implies that PKG-style retrieval is most appropriate in settings where (i) the corpus is relatively stable or updated incrementally, and (ii) the system serves many queries per build. For researchers, this implies that comparisons should include both accuracy and amortized costs, since build-time overhead is often negligible compared to repeated inference-time improvements at scale.

For both researchers and practitioners, the main takeaway is that structured retrieval (such as PKG) improves the quality of retrieved context relative to lexical baselines, but robust gains depend on careful selection and on mechanisms that prevent retrieval-induced regressions. The most promising path forward is to couple fine-grained retrieval with selection methods that are aligned with semantic correctness, while making retrieval conditional and budget-aware to preserve robustness across topics and models.

\subsection{Open Gaps and Directions}
\label{sec:disc-gaps}

Our analyses surface several gaps that future work can build on.
(1) \emph{Adaptive granularity:} choosing function- vs.\ block-level retrieval is currently static; developing query-adaptive policies could better match precision--recall needs across problem types.
(2) \emph{Robust text structuring:} the text-centric pipeline depends on converting tutorials into structured JSON; improving schema design and robustness of structuring under noisy or diverse prose remains open.
(3) \emph{Selection beyond retrieval:} post-generation reranking improves outcomes but is decoupled from retrieval; integrating retrieval-time uncertainty signals with candidate selection is an open design space.
(4) \emph{Cost-aware retrieval:} storage and preprocessing costs can dominate end-to-end feasibility; more cost-effective graph construction, compression, and caching strategies are needed.
These directions follow directly from the trade-offs observed in our topic/error/cost analyses and aim to make context retrieval for code generation both more reliable and more practical.

\section{Threats to Validity} \label{sec:threats}

Internal threats relates to the selection of the datasets and models. We have selected widely used datasets for code and tutorials, and models that are used in various code-intelligence studies. 
The embedder model was the top model on HuggingFace at the time of running the experiments. As the models used are considered across all experiments, the effects on the results are the same and therefore, mitigates any threats to the reliability of the results.

Though we have conducted experiments on no-RAG and RAG approaches and compared our work using open and closed source models, they are also conducted on Python and English tutorials. 
Thus the results are limited to these languages and more experiments are needed to evaluate the effects on other languages. 
We should also note that the scope of the results currently is limited to the dataset used to build the knowledge graph. Therefore, if other domains in Python or highly specific projects are considered, the results might not be directly applicable. In such cases, we recommend that a new knowledge graph be developed based on the context of the project. 

We anticipate minimal threats to the validity of the conducted tests. Though we used pass@1, we use a deterministic approach and therefore there is no need to conduct statistical tests and the results are directly comparable. 
Similarly, we provided various analysis on the topics and discussed the details from different angles, alleviating the conclusion validity of the experiments and results. 

\section{Conclusion}
\label{sec:conclusion}

We presented Programming Knowledge Graph, a novel approach for retrieval-augmented code generation, being both code-centric and text-centric. 
Our results lead to key findings.  
Retrieval is not uniformly beneficial for code generation. Structuring the retrieval space considering granularity improves reliability. Enhancing retrieval alone is insufficient and candidate selection with reranking is a primary driver of gains. 
The effectiveness of RAG approaches vary by topic and error mode; retrieval and reranking benefits are uneven across domains.
These findings suggest that retrieval-augmented code generation should be treated as a coupled problem of (i) constructing and indexing task-relevant artifacts at appropriate granularity, and (ii) robustly selecting among candidates in the presence of retrieval noise. Practically, this implies that future systems should emphasize retrieval gating, compute-parity evaluation, and rerankers that better correlate with semantic correctness. Promising directions include learning rerankers with execution-aware signals, incorporating lightweight specification checks for mathematically constrained topics, and developing adaptive retrieval policies that account for model capacity and topic characteristics. 
\bibliographystyle{ACM-Reference-Format}
\bibliography{references}

@inproceedings{wang-etal-2025-coderag,
  title     = {CodeRAG-Bench: Can Retrieval Augment Code Generation?},
  author    = {Wang, Zora Zhiruo and Asai, Akari and Yu, Xinyan Velocity and Xu, Frank F. and Xie, Yiqing and Neubig, Graham and Fried, Daniel},
  editor    = {Chiruzzo, Luis and Ritter, Alan and Wang, Lu},
  booktitle = {Findings of the Association for Computational Linguistics: NAACL 2025},
  month     = apr,
  year      = {2025},
  address   = {Albuquerque, New Mexico},
  publisher = {Association for Computational Linguistics},
  url       = {https://aclanthology.org/2025.findings-naacl.176/},
  doi       = {10.18653/v1/2025.findings-naacl.176},
  pages     = {3199--3214},
  isbn      = {979-8-89176-195-7}
}

@inproceedings{zhang-etal-2025-cast,
  title     = {c{AST}: Enhancing Code Retrieval-Augmented Generation with Structural Chunking via Abstract Syntax Tree},
  author    = {Zhang, Yilin and Zhao, Xinran and Wang, Zora Zhiruo and Yang, Chenyang and Wei, Jiayi and Wu, Tongshuang},
  editor    = {Christodoulopoulos, Christos and Chakraborty, Tanmoy and Rose, Carolyn and Peng, Violet},
  booktitle = {Findings of the Association for Computational Linguistics: EMNLP 2025},
  month     = nov,
  year      = {2025},
  address   = {Suzhou, China},
  publisher = {Association for Computational Linguistics},
  url       = {https://aclanthology.org/2025.findings-emnlp.430/},
  doi       = {10.18653/v1/2025.findings-emnlp.430},
  pages     = {8106--8116}
}

@inproceedings{sun-etal-2025-enhancing-retrieval,
  title     = {Enhancing Retrieval-Augmented Generation via Evidence Tree Search},
  author    = {Sun, Hao and Cai, Hengyi and Li, Yuchen and Fan, Xuanbo and Wei, Xiaochi and Wang, Shuaiqiang and Zhang, Yan and Yin, Dawei},
  editor    = {Che, Wanxiang and Nabende, Joyce and Shutova, Ekaterina and Pilehvar, Mohammad Taher},
  booktitle = {Proceedings of the 63rd Annual Meeting of the Association for Computational Linguistics (Volume 1: Long Papers)},
  month     = jul,
  year      = {2025},
  address   = {Vienna, Austria},
  publisher = {Association for Computational Linguistics},
  url       = {https://aclanthology.org/2025.acl-long.1175/},
  doi       = {10.18653/v1/2025.acl-long.1175},
  pages     = {24116--24127},
  isbn      = {979-8-89176-251-0}
}

@inproceedings{hagstrom-etal-2025-reality,
  title     = {A Reality Check on Context Utilisation for Retrieval-Augmented Generation},
  author    = {Hagstr{\"o}m, Lovisa and Marjanovic, Sara Vera and Yu, Haeun and Arora, Arnav and Lioma, Christina and Maistro, Maria and Atanasova, Pepa and Augenstein, Isabelle},
  editor    = {Che, Wanxiang and Nabende, Joyce and Shutova, Ekaterina and Pilehvar, Mohammad Taher},
  booktitle = {Proceedings of the 63rd Annual Meeting of the Association for Computational Linguistics (Volume 1: Long Papers)},
  month     = jul,
  year      = {2025},
  address   = {Vienna, Austria},
  publisher = {Association for Computational Linguistics},
  url       = {https://aclanthology.org/2025.acl-long.968/},
  doi       = {10.18653/v1/2025.acl-long.968},
  pages     = {19691--19730},
  isbn      = {979-8-89176-251-0}
}

@misc{pai2025codocbench,
  title        = {CoDocBench: A Dataset for Code-Documentation Alignment in Software Maintenance},
  author       = {Pai, Kunal and Devanbu, Premkumar and Ahmed, Toufique},
  year         = {2025},
  eprint       = {2502.00519},
  archivePrefix= {arXiv},
  primaryClass = {cs.SE},
  doi          = {10.48550/arXiv.2502.00519},
  url          = {https://arxiv.org/abs/2502.00519},
  note         = {Accepted at the 2025 IEEE/ACM 22nd International Conference on Mining Software Repositories (MSR) -- Data and Tool Showcase Track}
}

@misc{chirkova2025provence,
  title        = {Provence: efficient and robust context pruning for retrieval-augmented generation},
  author       = {Chirkova, Nadezhda and Formal, Thibault and Nikoulina, Vassilina and Clinchant, St{\'e}phane},
  year         = {2025},
  eprint       = {2501.16214},
  archivePrefix= {arXiv},
  primaryClass = {cs.CL},
  doi          = {10.48550/arXiv.2501.16214},
  url          = {https://arxiv.org/abs/2501.16214},
  note         = {Accepted to ICLR 2025}
}

@misc{sorokin2025reinfRerank,
  title        = {Iterative Self-Training for Code Generation via Reinforced Re-Ranking},
  author       = {Sorokin, Nikita and Sedykh, Ivan and Malykh, Valentin},
  year         = {2025},
  eprint       = {2504.09643},
  archivePrefix= {arXiv},
  primaryClass = {cs.CL},
  doi          = {10.48550/arXiv.2504.09643},
  url          = {https://arxiv.org/abs/2504.09643},
  note         = {Published at ECIR 2025; related DOI: 10.1007/978-3-031-88714-7\_21}
}

@misc{jiang2025ras,
  title        = {A Survey on Retrieval And Structuring Augmented Generation with Large Language Models},
  author       = {Jiang, Pengcheng and Ouyang, Siru and Jiao, Yizhu and Zhong, Ming and Tian, Runchu and Han, Jiawei},
  year         = {2025},
  eprint       = {2509.10697},
  archivePrefix= {arXiv},
  primaryClass = {cs.CL},
  doi          = {10.48550/arXiv.2509.10697},
  url          = {https://arxiv.org/abs/2509.10697}
}

@inproceedings{zhong-etal-2025-mix,
  title     = {Mix-of-Granularity: Optimize the Chunking Granularity for Retrieval-Augmented Generation},
  author    = {Zhong, Zijie and Liu, Hanwen and Cui, Xiaoya and Zhang, Xiaofan and Qin, Zengchang},
  editor    = {Rambow, Owen and Wanner, Leo and Apidianaki, Marianna and Al-Khalifa, Hend and Di Eugenio, Barbara and Schockaert, Steven},
  booktitle = {Proceedings of the 31st International Conference on Computational Linguistics},
  month     = jan,
  year      = {2025},
  address   = {Abu Dhabi, UAE},
  publisher = {Association for Computational Linguistics},
  pages     = {5756--5774},
  url       = {https://aclanthology.org/2025.coling-main.384/}
}

@misc{gu2025retrieve,
  title         = {What to Retrieve for Effective Retrieval-Augmented Code Generation? An Empirical Study and Beyond},
  author        = {Wenchao Gu and Juntao Chen and Yanlin Wang and Tianyue Jiang and Xingzhe Li and Mingwei Liu and Xilin Liu and Yuchi Ma and Zibin Zheng},
  year          = {2025},
  eprint        = {2503.20589},
  archivePrefix = {arXiv},
  primaryClass  = {cs.SE},
  url           = {https://arxiv.org/abs/2503.20589}
}

@misc{zhang2025graphragsurvey,
  title         = {A Survey of Graph Retrieval-Augmented Generation for Customized Large Language Models},
  author        = {Qinggang Zhang and Shengyuan Chen and Yuanchen Bei and Zheng Yuan and Huachi Zhou and Zijin Hong and Hao Chen and Yilin Xiao and Chuang Zhou and Junnan Dong and Yi Chang and Xiao Huang},
  year          = {2025},
  eprint        = {2501.13958},
  archivePrefix = {arXiv},
  primaryClass  = {cs.CL},
  url           = {https://arxiv.org/abs/2501.13958}
}

@misc{cao2025codersa,
  title         = {Pragmatic Reasoning improves {LLM} Code Generation},
  author        = {Zhuchen Cao and Sven Apel and Adish Singla and Vera Demberg},
  year          = {2025},
  eprint        = {2502.15835},
  archivePrefix = {arXiv},
  primaryClass  = {cs.CL},
  doi           = {10.48550/arXiv.2502.15835},
  url           = {https://arxiv.org/abs/2502.15835}
}

@inproceedings{jia2025structrag,
  title={StructRAG: Structure-Aware RAG Framework with Scholarly Knowledge Graph for Diverse Question Answering},
  author={Jia, Runsong and Zhang, Bowen and M{\'e}ndez, Sergio Jos{\'e} Rodr{\'\i}guez and Omran, Pouya G},
  booktitle={Companion Proceedings of the ACM on Web Conference 2025},
  pages={2567--2573},
  year={2025}
}

@article{zhu2025kg2rag,
  title={Knowledge graph-guided retrieval augmented generation},
  author={Zhu, Xiangrong and Xie, Yuexiang and Liu, Yi and Li, Yaliang and Hu, Wei},
  journal={arXiv preprint arXiv:2502.06864},
  year={2025}
}

@misc{hu2025grag,
      title={GRAG: Graph Retrieval-Augmented Generation}, 
      author={Yuntong Hu and Zhihan Lei and Zheng Zhang and Bo Pan and Chen Ling and Liang Zhao},
      year={2025},
      eprint={2405.16506},
      archivePrefix={arXiv},
      primaryClass={cs.LG},
      url={https://arxiv.org/abs/2405.16506}, 
}

@article{liang2025geographrag,
  title={GeoGraphRAG: A graph-based retrieval-augmented generation approach for empowering large language models in automated geospatial modeling},
  author={Liang, Jianyuan and Hou, Shuyang and Jiao, Haoyue and Qing, Yaxian and Zhao, Anqi and Shen, Zhangxiao and Xiang, Longgang and Wu, Huayi},
  journal={International Journal of Applied Earth Observation and Geoinformation},
  volume={142},
  pages={104712},
  year={2025},
  publisher={Elsevier}
}

@inproceedings{yang2025eventrag,
  title={EventRAG: Enhancing LLM Generation with Event Knowledge Graphs},
  author={Yang, Zairun and Wang, Yilin and Shi, Zhengyan and Yao, Yuan and Liang, Lei and Ding, Keyan and Yilmaz, Emine and Chen, Huajun and Zhang, Qiang},
  booktitle={Proceedings of the 63rd Annual Meeting of the Association for Computational Linguistics (Volume 1: Long Papers)},
  pages={16967--16979},
  year={2025}
}

@article{lin2025refrag,
  title={Refrag: Rethinking rag based decoding},
  author={Lin, Xiaoqiang and Ghosh, Aritra and Low, Bryan Kian Hsiang and Shrivastava, Anshumali and Mohan, Vijai},
  journal={arXiv preprint arXiv:2509.01092},
  year={2025}
}

@article{zhong2025semrag,
  title={SemRAG: Semantic Knowledge-Augmented RAG for Improved Question-Answering},
  author={Zhong, Kezhen and Suleiman, Basem and Erradi, Abdelkarim and Chen, Shijing},
  journal={arXiv preprint arXiv:2507.21110},
  year={2025}
}

@article{tao2025retrieval,
  title={Retrieval-Augmented Code Generation: A Survey with Focus on Repository-Level Approaches},
  author={Tao, Yicheng and Qin, Yao and Liu, Yepang},
  journal={arXiv e-prints},
  pages={arXiv--2510},
  year={2025}
}

@inproceedings{su-etal-2024-evor,
  title     = {EVOR: Evolving Retrieval for Code Generation},
  author    = {Su, Hongjin and Jiang, Shuyang and Lai, Yuhang and Wu, Haoyuan and Shi, Boao and Liu, Che and Liu, Qian and Yu, Tao},
  booktitle = {Findings of the Association for Computational Linguistics: EMNLP 2024},
  year      = {2024},
  publisher = {Association for Computational Linguistics},
  pages     = {2538--2554},
  url       = {https://aclanthology.org/2024.findings-emnlp.143/}
}

@misc{yang2025enhancingrepositorylevelsoftwarerepair,
      title={Enhancing repository-level software repair via repository-aware knowledge graphs}, 
      author={Boyang Yang and Jiadong Ren and Shunfu Jin and Yang Liu and Feng Liu and Bach Le and Haoye Tian},
      year={2025},
      eprint={2503.21710},
      archivePrefix={arXiv},
      primaryClass={cs.SE},
      url={https://arxiv.org/abs/2503.21710}, 
}

@misc{chen2025prometheusunifiedknowledgegraphs,
      title={Prometheus: Unified Knowledge Graphs for Issue Resolution in Multilingual Codebases}, 
      author={Zimin Chen and Yue Pan and Siyu Lu and Jiayi Xu and Claire Le Goues and Martin Monperrus and He Ye},
      year={2025},
      eprint={2507.19942},
      archivePrefix={arXiv},
      primaryClass={cs.SE},
      url={https://arxiv.org/abs/2507.19942}, 
}

@inproceedings{liu2025codexgraph,
  title={Codexgraph: Bridging large language models and code repositories via code graph databases},
  author={Liu, Xiangyan and Lan, Bo and Hu, Zhiyuan and Liu, Yang and Zhang, Zhicheng and Wang, Fei and Shieh, Michael Qizhe and Zhou, Wenmeng},
  booktitle={Proceedings of the 2025 Conference of the Nations of the Americas Chapter of the Association for Computational Linguistics: Human Language Technologies (Volume 1: Long Papers)},
  pages={142--160},
  year={2025}
}

@misc{ma2025alibabalingmaagentimprovingautomated,
      title={Alibaba LingmaAgent: Improving Automated Issue Resolution via Comprehensive Repository Exploration}, 
      author={Yingwei Ma and Qingping Yang and Rongyu Cao and Binhua Li and Fei Huang and Yongbin Li},
      year={2025},
      eprint={2406.01422},
      archivePrefix={arXiv},
      primaryClass={cs.SE},
      url={https://arxiv.org/abs/2406.01422}, 
}

@misc{ouyang2025repographenhancingaisoftware,
      title={RepoGraph: Enhancing AI Software Engineering with Repository-level Code Graph}, 
      author={Siru Ouyang and Wenhao Yu and Kaixin Ma and Zilin Xiao and Zhihan Zhang and Mengzhao Jia and Jiawei Han and Hongming Zhang and Dong Yu},
      year={2025},
      eprint={2410.14684},
      archivePrefix={arXiv},
      primaryClass={cs.SE},
      url={https://arxiv.org/abs/2410.14684}, 
}

@article{zhou2022docprompting,
  title={Docprompting: Generating code by retrieving the docs},
  author={Zhou, Shuyan and Alon, Uri and Xu, Frank F and Wang, Zhiruo and Jiang, Zhengbao and Neubig, Graham},
  journal={arXiv preprint arXiv: 2207.05987},
  year={2022}
}

@article{zhang2023repocoder,
  title={Repocoder: Repository-level code completion through iterative retrieval and generation},
  author={Zhang, Fengji and Chen, Bei and Zhang, Yue and Keung, Jacky and Liu, Jin and Zan, Daoguang and Mao, Yi and Lou, Jian-Guang and Chen, Weizhu},
  journal={arXiv preprint arXiv:2303.12570},
  year={2023}
}

@inproceedings{gutierrez2024hipporag,
  title={Hipporag: Neurobiologically inspired long-term memory for large language models},
  author={Guti{\'e}rrez, Bernal Jim{\'e}nez and Shu, Yiheng and Gu, Yu and Yasunaga, Michihiro and Su, Yu},
  booktitle={The Thirty-eighth Annual Conference on Neural Information Processing Systems},
  year={2024}
}

@article{roziere2023code,
  title={Code llama: Open foundation models for code},
  author={Roziere, Baptiste and Gehring, Jonas and Gloeckle, Fabian and Sootla, Sten and Gat, Itai and Tan, Xiaoqing Ellen and Adi, Yossi and Liu, Jingyu and Remez, Tal and Rapin, J{\'e}r{\'e}my and others},
  journal={arXiv preprint arXiv:2308.12950},
  year={2023}
}

@article{huang2023agentcoder,
  title={Agentcoder: Multi-agent-based code generation with iterative testing and optimisation},
  author={Huang, Dong and Bu, Qingwen and Zhang, Jie M and Luck, Michael and Cui, Heming},
  journal={arXiv preprint arXiv:2312.13010},
  year={2023}
}

@article{chen2021evaluating,
  title={Evaluating large language models trained on code},
  author={Chen, Mark and Tworek, Jerry and Jun, Heewoo and Yuan, Qiming and Pinto, Henrique Ponde De Oliveira and Kaplan, Jared and Edwards, Harri and Burda, Yuri and Joseph, Nicholas and Brockman, Greg and others},
  journal={arXiv preprint arXiv:2107.03374},
  year={2021}
}

@article{austin2021mbpp,
  title={Program synthesis with large language models},
  author={Austin, Jacob and Odena, Augustus and Nye, Maxwell and Bosma, Maarten and Michalewski, Henryk and Dohan, David and Jiang, Ellen and Cai, Carrie and Terry, Michael and Le, Quoc and others},
  journal={arXiv preprint arXiv:2108.07732},
  year={2021}
}

@inproceedings{parvez2021retrieval,
  title={Retrieval Augmented Code Generation and Summarization},
  author={Parvez, Md Rizwan and Ahmad, Wasi and Chakraborty, Saikat and Ray, Baishakhi and Chang, Kai-Wei},
  booktitle={Findings of the Association for Computational Linguistics: EMNLP 2021},
  pages={2719--2734},
  year={2021}
}

@article{gao2023retrieval,
  title={Retrieval-augmented generation for large language models: A survey},
  author={Gao, Yunfan and Xiong, Yun and Gao, Xinyu and Jia, Kangxiang and Pan, Jinliu and Bi, Yuxi and Dai, Yi and Sun, Jiawei and Wang, Haofen},
  journal={arXiv preprint arXiv:2312.10997},
  year={2023}
}

@inproceedings{jiang2023active,
  title={Active Retrieval Augmented Generation},
  author={Jiang, Zhengbao and Xu, Frank F and Gao, Luyu and Sun, Zhiqing and Liu, Qian and Dwivedi-Yu, Jane and Yang, Yiming and Callan, Jamie and Neubig, Graham},
  booktitle={Proceedings of the 2023 Conference on Empirical Methods in Natural Language Processing},
  pages={7969--7992},
  year={2023}
}

@article{wang2024coderag,
  title={CodeRAG-Bench: Can Retrieval Augment Code Generation?},
  author={Wang, Zora Zhiruo and Asai, Akari and Yu, Xinyan Velocity and Xu, Frank F and Xie, Yiqing and Neubig, Graham and Fried, Daniel},
  journal={arXiv preprint arXiv:2406.14497},
  year={2024}
}

@article{lewis2020retrieval,
  title={Retrieval-augmented generation for knowledge-intensive nlp tasks},
  author={Lewis, Patrick and Perez, Ethan and Piktus, Aleksandra and Petroni, Fabio and Karpukhin, Vladimir and Goyal, Naman and K{\"u}ttler, Heinrich and Lewis, Mike and Yih, Wen-tau and Rockt{\"a}schel, Tim and others},
  journal={Advances in Neural Information Processing Systems},
  volume={33},
  pages={9459--9474},
  year={2020}
}

@article{team2024gemma,
  title={Gemma 2: Improving open language models at a practical size},
  author={Team, Gemma and Riviere, Morgane and Pathak, Shreya and Sessa, Pier Giuseppe and Hardin, Cassidy and Bhupatiraju, Surya and Hussenot, L{\'e}onard and Mesnard, Thomas and Shahriari, Bobak and Ram{\'e}, Alexandre and others},
  journal={arXiv preprint arXiv:2408.00118},
  year={2024}
}

@misc{nic2024pyalpaca,
  author = {Nicolas Mejia Petit},
  title = {Tested-143k-Python-Alpaca},
  year = 2024,
  url = {https://huggingface.co/datasets/Vezora/Tested-143k-Python-Alpaca},
  urldate = {2024-03-06}
}

@inproceedings{guu2020retrieval,
  title={Retrieval augmented language model pre-training},
  author={Guu, Kelvin and Lee, Kenton and Tung, Zora and Pasupat, Panupong and Chang, Mingwei},
  booktitle={International conference on machine learning},
  pages={3929--3938},
  year={2020},
  organization={PMLR}
}

@inproceedings{zhoudocprompting,
  title={DocPrompting: Generating Code by Retrieving the Docs},
  author={Zhou, Shuyan and Alon, Uri and Xu, Frank F and Jiang, Zhengbao and Neubig, Graham},
  booktitle={The Eleventh International Conference on Learning Representations},
  year={2022}
}

@article{dubey2024llama3,
  title={The llama 3 herd of models},
  author={Dubey, Abhimanyu and Jauhri, Abhinav and Pandey, Abhinav and Kadian, Abhishek and Al-Dahle, Ahmad and Letman, Aiesha and Mathur, Akhil and Schelten, Alan and Yang, Amy and Fan, Angela and others},
  journal={arXiv preprint arXiv:2407.21783},
  year={2024}
}

@article{zhu2024deepseek,
  title={DeepSeek-Coder-V2: Breaking the Barrier of Closed-Source Models in Code Intelligence},
  author={Zhu, Qihao and Guo, Daya and Shao, Zhihong and Yang, Dejian and Wang, Peiyi and Xu, Runxin and Wu, Y and Li, Yukun and Gao, Huazuo and Ma, Shirong and others},
  journal={arXiv preprint arXiv:2406.11931},
  year={2024}
}

@article{roziere2023codellama,
  title={Code llama: Open foundation models for code},
  author={Roziere, Baptiste and Gehring, Jonas and Gloeckle, Fabian and Sootla, Sten and Gat, Itai and Tan, Xiaoqing Ellen and Adi, Yossi and Liu, Jingyu and Sauvestre, Romain and Remez, Tal and others},
  journal={arXiv preprint arXiv:2308.12950},
  year={2023}
}

@article{lozhkov2024starcoder2,
  title={Starcoder 2 and the stack v2: The next generation},
  author={Lozhkov, Anton and Li, Raymond and Allal, Loubna Ben and Cassano, Federico and Lamy-Poirier, Joel and Tazi, Nouamane and Tang, Ao and Pykhtar, Dmytro and Liu, Jiawei and Wei, Yuxiang and others},
  journal={arXiv preprint arXiv:2402.19173},
  year={2024}
}

\appendix
\section{PKG Schema Inventory}
\label{app:pkg-schema}

This appendix lists the concrete node and edge types used to instantiate the abstract schema in Section~\ref{sec:pkg-generation}.

\subsection{Node types}
\label{app:pkg_node_types}
We use the following node types:
\begin{itemize}
    \item \textsc{Name}: a function identifier (string).
    \item \textsc{Impl}: a full function implementation (code).
    \item \textsc{Block}: a syntactic code block span inside a function (code).
    \item \textsc{PathValue}: a (path, value) leaf extracted from a JSON representation of tutorial content (text), following Section~\ref{sec:json_extraction}.
\end{itemize}

\subsection{Edge types}
\label{app:pkg_edge_types}
Edges capture identifier-to-content links and refinement (containment) relations:
\begin{itemize}
    \item \textsc{has\_impl}: \textsc{Name} $\rightarrow$ \textsc{Impl}.
    \item \textsc{has\_block}: \textsc{Impl} $\rightarrow$ \textsc{Block}.
    \item \textsc{parent}: \textsc{Block} $\rightarrow$ \textsc{Block} (syntactic block nesting).
    \item \textsc{json\_child}: \textsc{PathValue} $\rightarrow$ \textsc{PathValue} (JSON hierarchy when internal path nodes are materialized, see Section~\ref{sec:json_extraction}).
\end{itemize}

\section{Examples and Prompts for PKG Approach}

\subsection{Challenges in Retrieving Information from PKG}
\label{sec:challenges}

This section discusses scenarios where the PKG fails to retrieve accurate or relevant information. One notable challenge arises when the task requires domain-specific expertise. For example, if the task involves a specialized framework or project-specific code, the PKG must be populated with relevant data from the corresponding domain or project. Failures occur when queries target a graph that lacks such domain knowledge. Addressing this issue necessitates updating the graph with appropriate domain-specific information.

Through topic analysis, we identified that the PKG often struggles with certain problem categories, such as string manipulation. Experimental observations indicate that this challenge stems from the limitations of both the embedder model and the baseline model, which tend to prioritize semantic meaning over structural characteristics of strings.

Example Problem:
Write a Python function to convert lowercase characters to uppercase and vice versa, transforming inputs such as "Hello" into "hELLO" and "pYthon" into "PyTHON".

Challenges:
\begin{itemize}
    \item Embedding Model’s Semantic Bias:
    
    In RAG, the embedder retrieves content primarily based on semantic meaning rather than formatting or structural patterns. For example, it might interpret "Hello" as a greeting, ignoring the case transformation requirement.

    \item LLM’s Tokenization and Semantic Prioritization:

    LLMs tokenize text based on meaning rather than formatting. Consequently, tokens like "Hello" and "hello" are often treated identically, making tasks involving case transformations particularly challenging.
\end{itemize}

In summary, both RAG retrieval and LLM tokenization emphasize semantic understanding over structural or formatting details, complicating the handling of tasks like string manipulation. This limitation reduces the effectiveness of PKG-based approaches for such problem categories.

\subsection{CodeLLama7b}
\subsubsection{Prompts}

The prompts we have used for CodeLlama7B model is provided in Code \ref{code:codellama}:

\begin{lstlisting}[label={code:codellama}]
def codellama_prompt(problem,augmented_data=None):
    if augmented_data:
        prompt = f"""[INST] You are a python programmer. Solve the following problem:\n{problem} \n\nThe following code might be helpful:\n{augmented_data}\nIf helper section is useful, integrate their logic directly into the body of the main function, otherwise just ignore them. You MUST write your solution between [PYTHON] and [/PYTHON]. Your solution MUST be executable.[/INST]"""
        return prompt
    else:
        prompt = f"""[INST] You are a python programmer. Solve the following problem:\n{problem} \n\nPlease write the python solution inside [PYTHON] and [/PYTHON] tags.\n[/INST]"
        """
        return prompt
\end{lstlisting}

\begin{figure}[htbp]
    \centering
    \includegraphics[width=0.80\textwidth]{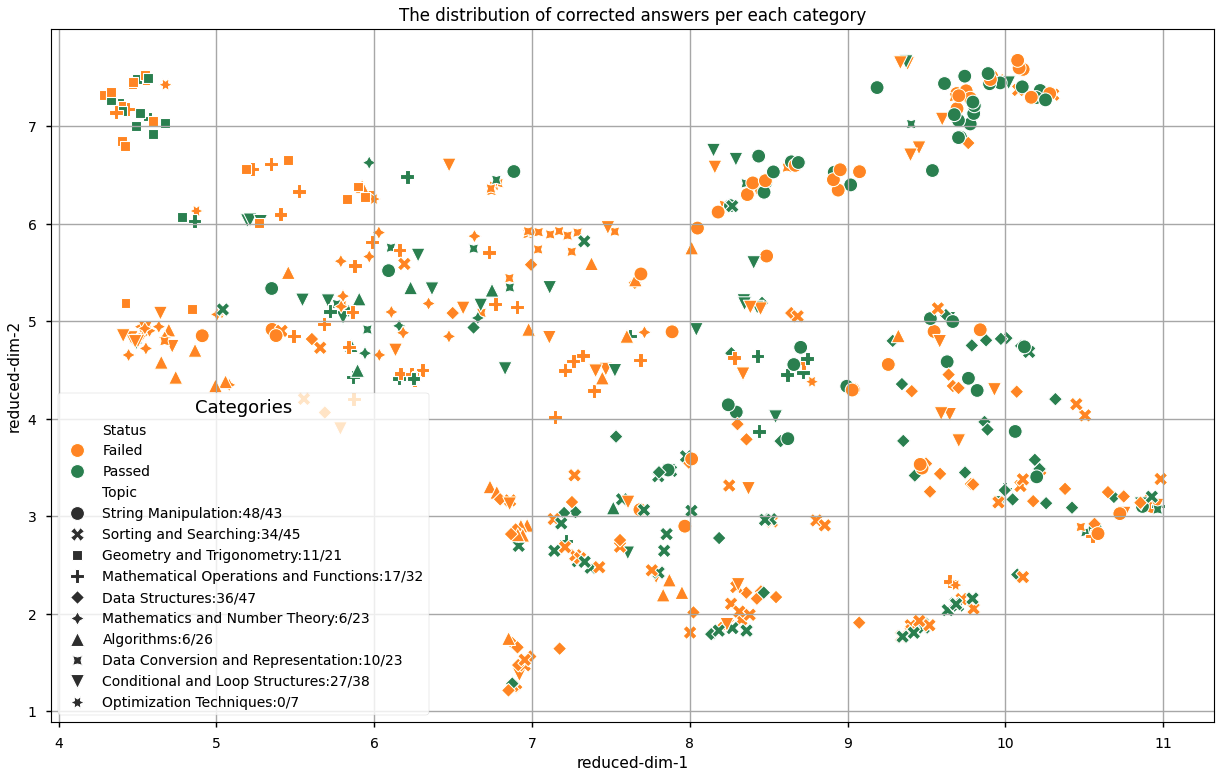}
    \Description{The distribution of MBPP solutions on each topic in NoRAG setting using CodeLLama7b.}
    \caption{The distribution of MBPP solutions on each topic in NoRAG setting using CodeLLama7b.}
    \label{fig:topic_based_accuracy_codellama7b_norag}
\end{figure}

\subsubsection{Topic-Based Accuracy distribution}
\begin{figure}[htbp]
    \centering
    \includegraphics[width=0.80\textwidth]{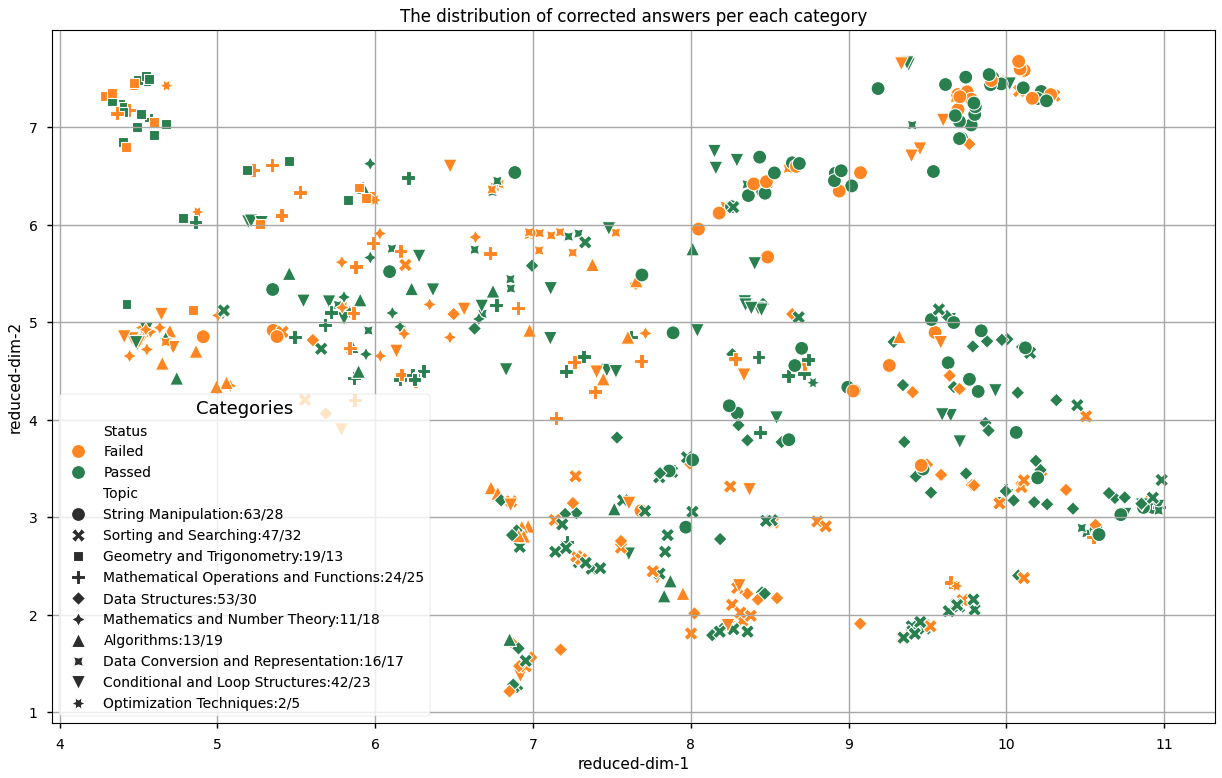}
    \caption{The distribution of MBPP solutions on each topic using our re-ranker using CodeLLama7b.}
    \label{fig:topic_based_accuracy_codellama7b_reranked}
    \Description{The distribution of MBPP solutions on each topic using our re-ranker using CodeLLama7b.}
\end{figure}

Figure \ref{fig:topic_based_accuracy_codellama7b_norag} illustrates the distribution of MBPP problems on a two-dimensional plot, where the embedding dimensions have been reduced to two for visualization purposes. The different problem topics are represented by distinct shapes, while the correctness of the solutions is indicated by color. Problems that were solved incorrectly are shown in orange, and those solved correctly are shown in green. The legend for each topic separates the total number of correct solutions from the incorrect ones using a slash ("/"). Figure \ref{fig:topic_based_accuracy_codellama7b_reranked} shows the distribution of correct and incorrect problems when we apply our approach.

\subsection{StarCoder2-7B}
\subsubsection{Prompts}

The prompts we have used for StarCoder2-7B model is provided in Code \ref{code:starcoder}:

\begin{lstlisting}[label={code:starcoder}]

def starcoder_prompt(problem,augmented_data=None):
    if augmented_data:
        prompt = f"""### Instruction
        You are a python programmer. Solve the following problem:\n{problem} \n\n The following code might be helpful:\n{augmented_data}\n. If they are useful, integrate their logic directly into the body of the main function, otherwise just ignore them.\n
        ### Response
        """
        return prompt
    else:
        prompt = f"""### Instruction
        You are a python programmer. Solve the following problem:\n{problem} \n\n
        ### Response
        """
        return prompt
\end{lstlisting}

\begin{figure}[htbp]
    \centering
    \includegraphics[width=0.80\textwidth]{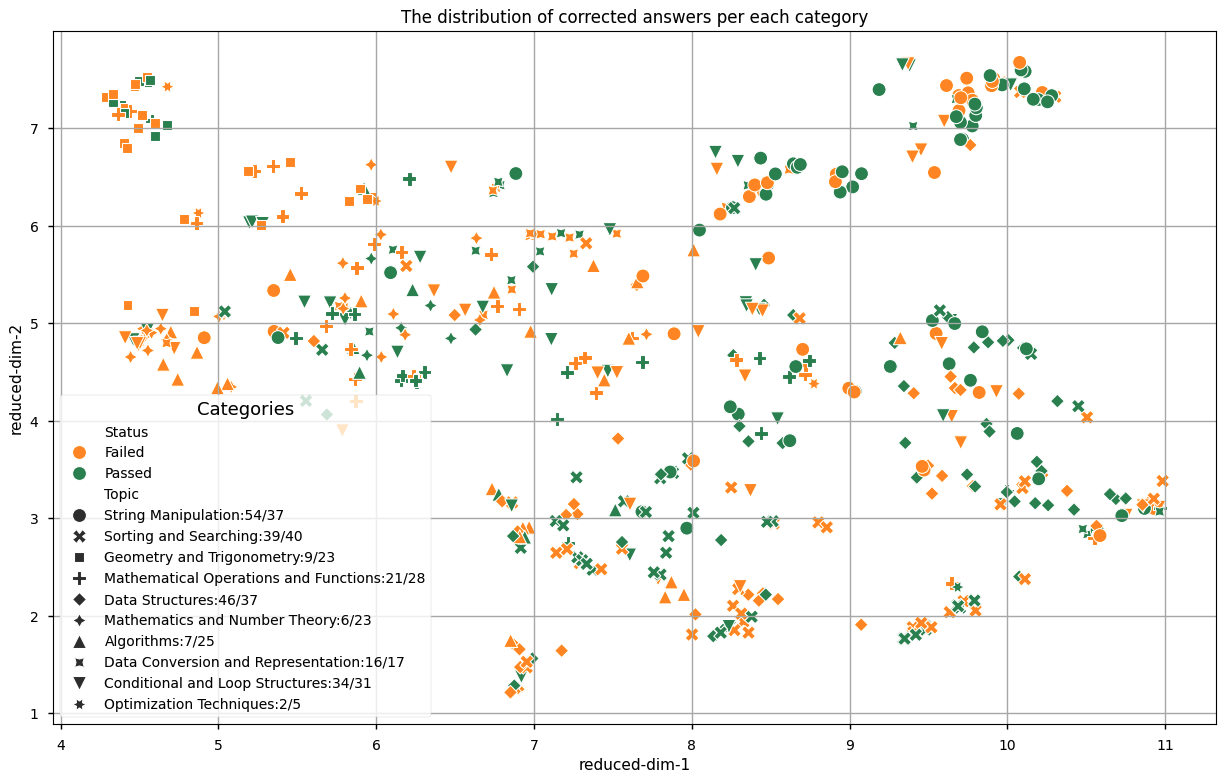}
    \caption{The distribution of MBPP solutions on each topic without RAG using StarCoder2-7B.}
    \label{fig:topic_based_accuracy_starcoder}
    \Description{The distribution of MBPP solutions on each topic without RAG using StarCoder2-7B.}
\end{figure}

\subsubsection{Topic-Based Accuracy Distribution}
Figure \ref{fig:topic_based_accuracy_starcoder} presents the distribution of MBPP problems on a two-dimensional plot, with the embedding dimensions reduced for visualization. Each problem topic is represented by a unique shape, while solution correctness is color-coded. Problems incorrectly solved by StarCoder2-7B are highlighted in orange, whereas correctly solved problems are shown in green. The legend for each topic indicates the total number of correct versus incorrect solutions using a "correct/incorrect" format.

\begin{figure}[htbp]
    \centering
    \includegraphics[width=0.80\textwidth]{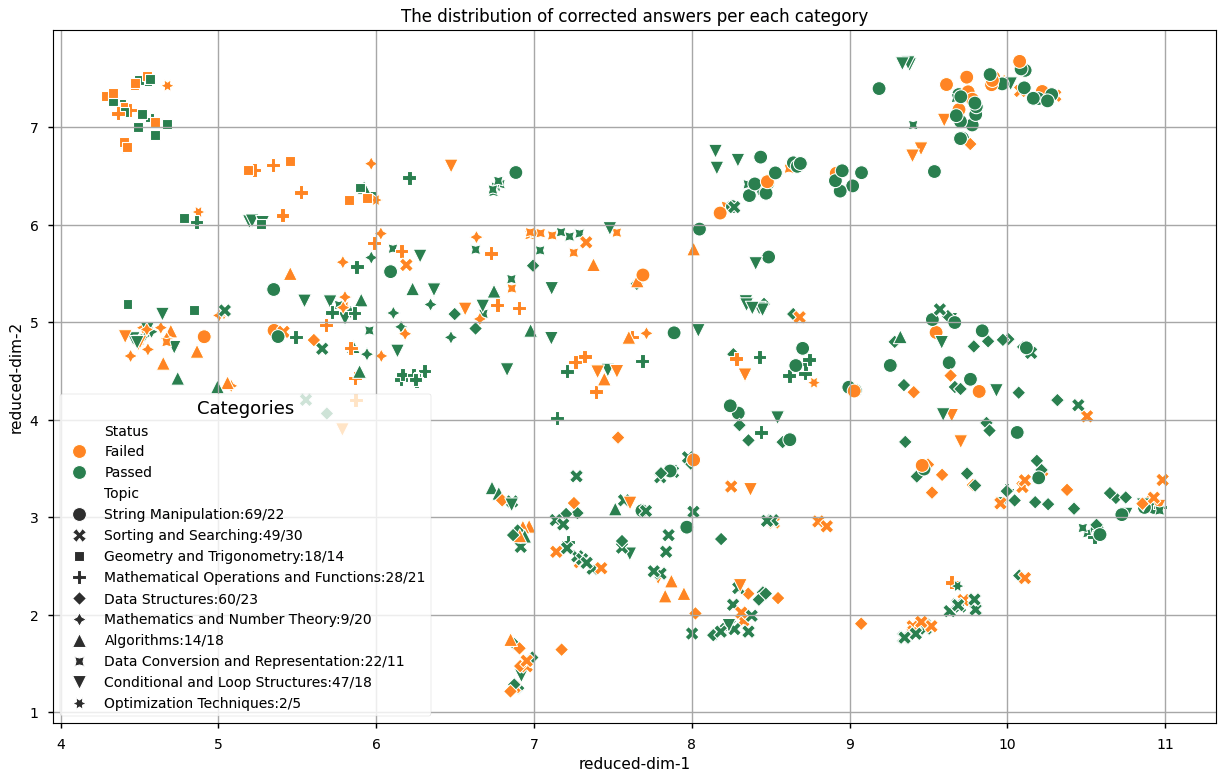}
    \Description{The distribution of MBPP solutions on each topic using our proposed re-ranker using StarCoder2-7B.}
    \caption{The distribution of MBPP solutions on each topic using our proposed re-ranker using StarCoder2-7B.}
    \label{fig:topic_based_accuracy_starcoder_reranked}
\end{figure}

Additionally, Figure \ref{fig:topic_based_accuracy_starcoder_reranked}
 visualizes the same distribution but reflects the accuracy after applying our proposed approach, showcasing improvements in solution correctness across topics.

\subsection{DeepSeek-Coder-7B}
\subsubsection{Prompts}

The prompts we have used for DeepSeek-Coder-7B model is provided in Code \ref{code:deepseek}:

\begin{lstlisting}[label={code:deepseek}]

def deepseek_prompt(problem,augmented_data=None):
    if augmented_data:
        prompt = f"""[INST] You are a python programmer. Solve the following problem:\n{problem} \n\n The following code might be helpful:\n{augmented_data}\n.If they are useful, integrate their logic directly into the body of the main function, otherwise just ignore them.\n[/INST]"""
        return prompt
    else:
        prompt = f"""[INST] You are a python programmer.  Solve the following problem: \n {problem} \n\n[/INST]"""
        return prompt
\end{lstlisting}

\subsubsection{Topic-Based Accuracy Distribution}

\begin{figure}[htbp]
    \centering
    \includegraphics[width=0.80\textwidth]{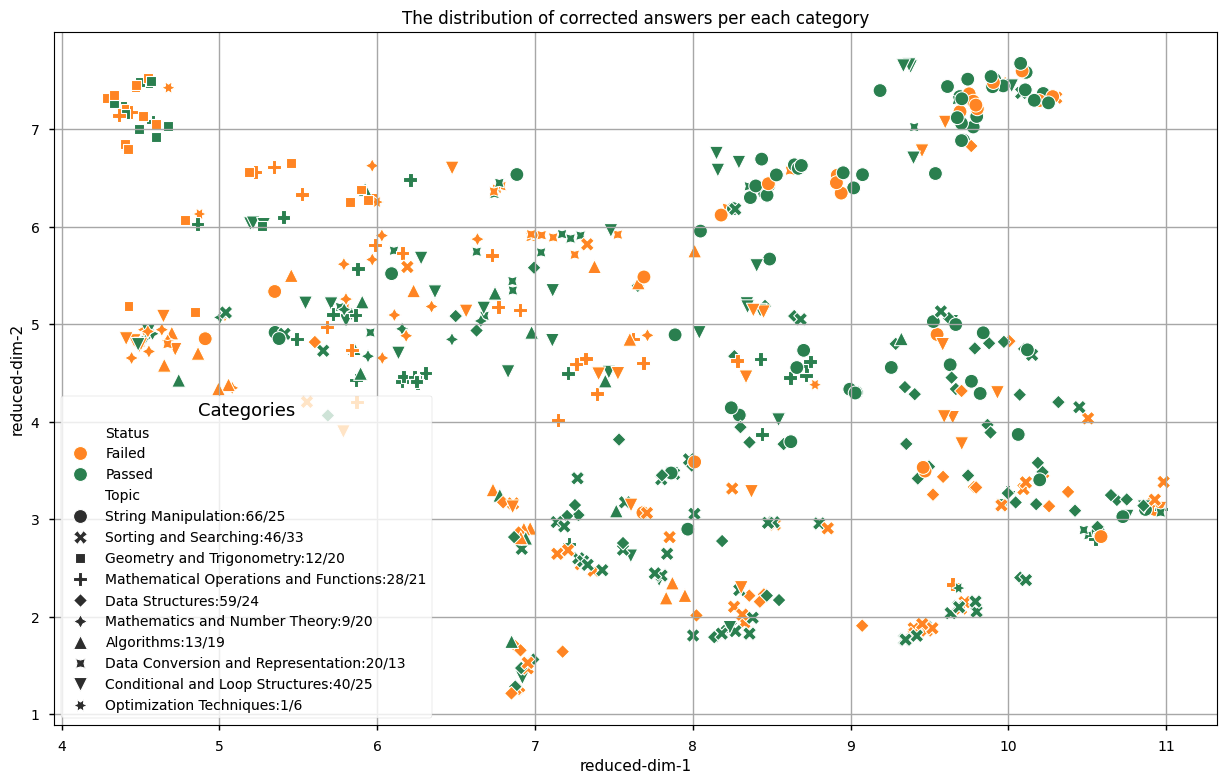}
    \caption{The distribution of MBPP solutions on each topic in NoRAG setting using DeepSeek-Coder-7B.}
    \label{fig:topic_based_accuracy_deepseek_norag}
    \Description{The distribution of MBPP solutions on each topic in NoRAG setting using DeepSeek-Coder-7B.}
\end{figure}

Figure \ref{fig:topic_based_accuracy_deepseek_norag} displays the distribution of problems from the MBPP dataset in a two-dimensional plot, achieved by reducing the embedding dimensions for improved visualization. Each distinct shape in the plot corresponds to a specific problem topic, while the correctness of the solutions is indicated by color coding. Problems that were solved incorrectly are represented in orange, whereas those that were solved correctly are shown in green. The legend accompanying each topic delineates the total number of correct solutions from the incorrect ones, separated with a slash ("/").

\begin{figure}[htbp]
    \centering
    \includegraphics[width=0.80\textwidth]{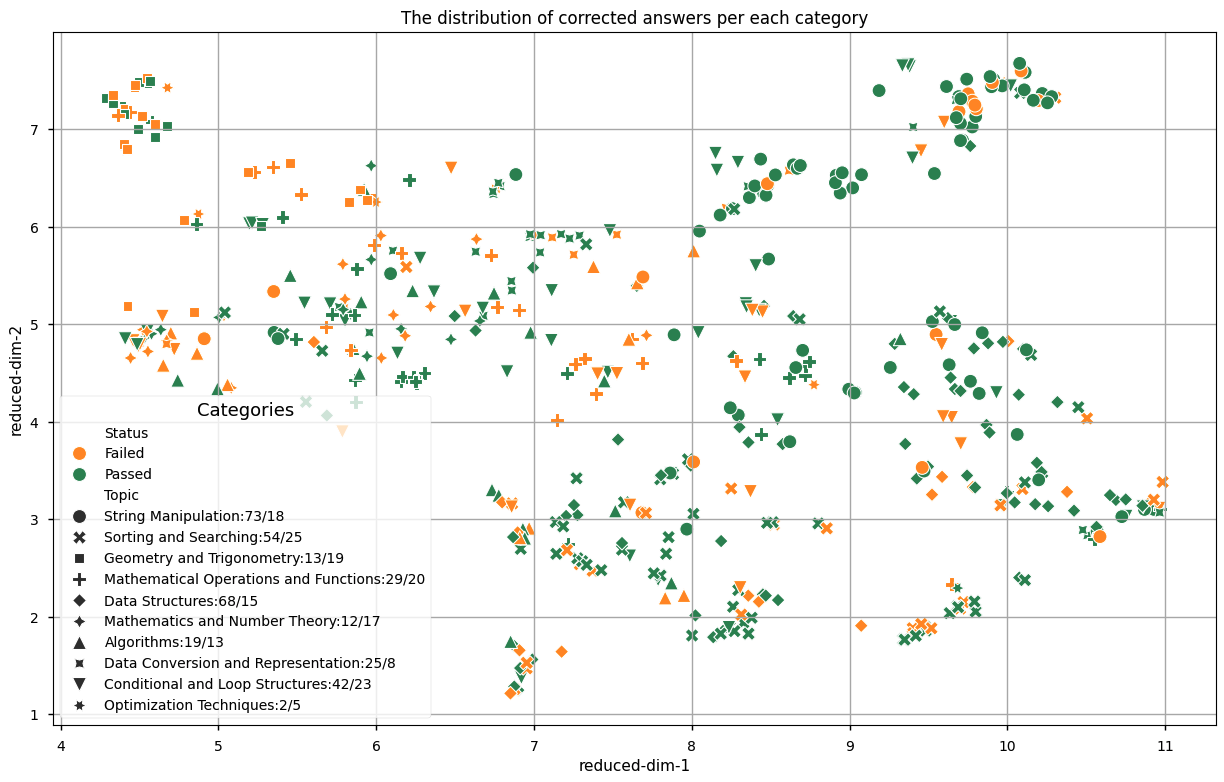}
    \caption{The distribution of MBPP solutions on each topic using our reranker using DeepSeek-Coder-7B.}
    \label{fig:topic_based_accuracy_deepseek_reranked}
    \Description{The distribution of MBPP solutions on each topic using our reranker using DeepSeek-Coder-7B.}
\end{figure}

In addition, Figure \ref{fig:topic_based_accuracy_deepseek_reranked} presents a similar distribution of problems, highlighting the outcomes after applying our novel approach. This figure further distinguishes between correct and incorrect solutions, allowing for a comparative analysis of the effectiveness of our method.

\subsection{Examples}
\label{appendix:examples}
In this section, we present two selected samples from the HumanEval benchmark. We provide the responses generated by StarCoder-2-7B and DeepSeek-Coder-7B models. Each model's output is displayed in two scenarios: first, without using RAG, and second, utilizing our PKG approach. These examples illustrate how incorporating additional context can enhance the models' ability to solve complex problems more effectively.

HumanEval problem 159, solved by Starcoder2-7B without RAG (Failed):

\begin{lstlisting}

def eat(number, need, remaining):
    """
    You're a hungry rabbit, and you already have eaten a certain number of carrots,
    but now you need to eat more carrots to complete the day's meals.
    you should return an array of [ total number of eaten carrots after your meals,
                                    the number of carrots left after your meals ]
    if there are not enough remaining carrots, you will eat all remaining carrots, but will still be hungry.
    
    Example:
    * eat(5, 6, 10) -> [11, 4]
    * eat(4, 8, 9) -> [12, 1]
    * eat(1, 10, 10) -> [11, 0]
    * eat(2, 11, 5) -> [7, 0]
    
    Variables:
    @number : integer
        the number of carrots that you have eaten.
    @need : integer
        the number of carrots that you need to eat.
    @remaining : integer
        the number of remaining carrots thet exist in stock
    
    Constrain:
    * 0 <= number <= 1000
    * 0 <= need <= 1000
    * 0 <= remaining <= 1000

    Have fun :)
    """
    return [number + need, remaining - need]
\end{lstlisting}

HumanEval problem 159, solved by Starcoder2-7B using PKG (Passed):

\begin{lstlisting}

### Instruction
#You are a python expert programmer. Implement a python function to #solve the following problem: 
 
def eat(number, need, remaining):
    """
    You're a hungry rabbit, and you already have eaten a certain number of carrots,
    but now you need to eat more carrots to complete the day's meals.
    you should return an array of [ total number of eaten carrots after your meals,
                                    the number of carrots left after your meals ]
    if there are not enough remaining carrots, you will eat all remaining carrots, but will still be hungry.
    
    Example:
    * eat(5, 6, 10) -> [11, 4]
    * eat(4, 8, 9) -> [12, 1]
    * eat(1, 10, 10) -> [11, 0]
    * eat(2, 11, 5) -> [7, 0]
    
    Variables:
    @number : integer
        the number of carrots that you have eaten.
    @need : integer
        the number of carrots that you need to eat.
    @remaining : integer
        the number of remaining carrots thet exist in stock
    
    Constrain:
    * 0 <= number <= 1000
    * 0 <= need <= 1000
    * 0 <= remaining <= 1000

    Have fun :)
    """
 

#The following helper code might be helpful:
#helper code 1:
    postEatWeight = totalWeight - amountEaten
    combinedWeight = postEatWeight / 7
    largerBoxPostEat = combinedWeight * 4
    smallerBoxPostEat = combinedWeight * 3
    largerBox = largerBoxPostEat + amountEaten
    smallerBox = smallerBoxPostEat
    return (round(largerBox), round(smallerBox))
#End of helper section.
#Write your solution inside [PYTHON] and [/PYTHON] tags. The solution #should be executable and do NOT call helper functions. If they are #useful, integrate their logic directly into the body of the main #function, otherwise just ignore them.
### Response
#Here is the Python function to solve the problem:


def eat(number, need, remaining):
    if remaining >= need:
        return [number + need, remaining - need]
    else:
        return [number + remaining, 0]

\end{lstlisting}

HumanEval problem 90, solved by DeepSeek-Coder2 without RAG (Failed):

\begin{lstlisting}
def next_smallest(lst):
    """
    You are given a list of integers.
    Write a function next_smallest() that returns the 2nd smallest element of the list.
    Return None if there is no such element.
    
    next_smallest([1, 2, 3, 4, 5]) == 2
    next_smallest([5, 1, 4, 3, 2]) == 2
    next_smallest([]) == None
    next_smallest([1, 1]) == None
    """
    if len(lst) < 2:
        return None
    else:
        lst.sort()
        return lst[1]
        
\end{lstlisting}

HumanEval problem 90, solved by DeepSeek-Coder2 using PKG (Passed):

\begin{lstlisting}
### Instruction:
#You are an expert python programmer. Solve the following PROBLEM in #python: 

def next_smallest(lst):
    """
    You are given a list of integers.
    Write a function next_smallest() that returns the 2nd smallest element of the list.
    Return None if there is no such element.
    
    next_smallest([1, 2, 3, 4, 5]) == 2
    next_smallest([5, 1, 4, 3, 2]) == 2
    next_smallest([]) == None
    next_smallest([1, 1]) == None
    """
 
 #The following helper code might be helpful:
 #helper code 1:
    if len(lst) < 2:
        return None
    smallest = float('inf')
    next_smallest = float('inf')
    for i in range(0, len(lst)):
        if lst[i] < smallest:
            next_smallest = smallest
            smallest = lst[i]
        elif lst[i] > smallest and lst[i] < next_smallest:
            next_smallest = lst[i]
    return None if next_smallest == float('inf') else next_smallest

#End of helper section.
#Please write the python solution inside [PYTHON] and [/PYTHON] tags.
### Response:

def next_smallest(lst):
    if len(lst) < 2:
        return None
    smallest = second_smallest = float('inf')
    for element in lst:
        if element < smallest:
            second_smallest = smallest
            smallest = element
        elif smallest < element < second_smallest:
            second_smallest = element
    return None if second_smallest == float('inf') else second_smallest

\end{lstlisting}

\end{document}